\documentstyle[11pt,aaspp4]{article}   

\slugcomment{Accepted by The Astronomical Journal}

\lefthead{Schneider et al.}
\righthead{SDSS Quasar Catalog II.}

\begin{document}

\title{The Sloan Digital Sky Survey Quasar Catalog~II. First Data Release
}

\author{
Donald~P.~Schneider,\altaffilmark{\ref{PennState}}
Xiaohui~Fan,\altaffilmark{\ref{Arizona}}
Patrick~B.~Hall,\altaffilmark{\ref{Princeton}}$^,$\altaffilmark{\ref{CChile}}
Sebastian~Jester,\altaffilmark{\ref{FNAL}}
Gordon~T.~Richards,\altaffilmark{\ref{PennState}}$^,$\altaffilmark{\ref{Princeton}}
Chris~Stoughton,\altaffilmark{\ref{FNAL}}
Michael~A.~Strauss,\altaffilmark{\ref{Princeton}}
Mark~SubbaRao,\altaffilmark{\ref{Chicago}}$^,$\altaffilmark{\ref{Adler}}
Daniel~E.~Vanden~Berk,\altaffilmark{\ref{Pittsburgh}}
Scott~F.~Anderson,\altaffilmark{\ref{UW}}
W.N.~Brandt,\altaffilmark{\ref{PennState}}
James~E.~Gunn,\altaffilmark{\ref{Princeton}}
Jim~Gray,\altaffilmark{\ref{MicroSoft}}
Jonathan~R.~Trump,\altaffilmark{\ref{PennState}}
Wolfgang~Voges,\altaffilmark{\ref{MPE}}
Brian~Yanny,\altaffilmark{\ref{FNAL}}
Neta~A.~Bahcall,\altaffilmark{\ref{Princeton}}
Michael~R.~Blanton,\altaffilmark{\ref{NYU}}
William~N.~Boroski,\altaffilmark{\ref{FNAL}}
J.~Brinkmann,\altaffilmark{\ref{APO}}
Robert~Brunner,\altaffilmark{\ref{Caltech}}
Scott~Burles,\altaffilmark{\ref{MIT}}
Francisco~J.~Castander,\altaffilmark{\ref{Barcelona}}
Mamoru~Doi,\altaffilmark{\ref{Tokyo2}}
Daniel~Eisenstein,\altaffilmark{\ref{Arizona}}
Joshua~A.~Frieman,\altaffilmark{\ref{Chicago}}
Masataka~Fukugita,\altaffilmark{\ref{CosJapan}}$^,$\altaffilmark{\ref{IAS}}
Timothy~M.~Heckman,\altaffilmark{\ref{JHU}}
G.S.~Hennessy,\altaffilmark{\ref{USNODC}}
\v Zeljko~Ivezi\'c,\altaffilmark{\ref{Princeton}}$^,$\altaffilmark{\ref{HNRus}}
Stephen~Kent,\altaffilmark{\ref{FNAL}}
Gillian~R.~Knapp,\altaffilmark{\ref{Princeton}}
Donald~Q.~Lamb,\altaffilmark{\ref{Chicago}}$^,$\altaffilmark{\ref{Fermi}}
Brian~C.~Lee,\altaffilmark{\ref{LBNL}}
Jon~Loveday,\altaffilmark{\ref{Sussex}}
Robert~H.~Lupton,\altaffilmark{\ref{Princeton}}
Bruce~Margon,\altaffilmark{\ref{STScI}}
Avery~Meiksin,\altaffilmark{\ref{Edinburgh}}
Jeffrey~A.~Munn,\altaffilmark{\ref{USNOAZ}}
Heidi~Jo~Newberg,\altaffilmark{\ref{RPI}}
R.C.~Nichol,\altaffilmark{\ref{CMU}}
Martin~Niederste-Ostholt,\altaffilmark{\ref{Princeton}}
Jeffrey~R.~Pier,\altaffilmark{\ref{USNOAZ}}
Michael~W.~Richmond,\altaffilmark{\ref{RIT}}
Constance~M.~Rockosi,\altaffilmark{\ref{UW}}
David~H.~Saxe,\altaffilmark{\ref{IAS}}
David~J.~Schlegel,\altaffilmark{\ref{Princeton}}
Alexander~S.~Szalay,\altaffilmark{\ref{JHU}}
Aniruddha~R.~Thakar,\altaffilmark{\ref{JHU}}
Alan~Uomoto,\altaffilmark{\ref{JHU}}
and
Donald~G.~York\altaffilmark{\ref{Chicago}}$^,$\altaffilmark{\ref{Fermi}}
}


\newcounter{address}
\setcounter{address}{1}
\altaffiltext{\theaddress}{Department of Astronomy and Astrophysics, The
   Pennsylvania State University, University Park, PA 16802.
\label{PennState}}
\addtocounter{address}{1}
\altaffiltext{\theaddress}{Steward Observatory, The University of Arizona,
   933~North Cherry Avenue, Tucson, AZ 85721.
\label{Arizona}}
\addtocounter{address}{1}
\altaffiltext{\theaddress}{Princeton University Observatory, Princeton,
   NJ 08544.
\label{Princeton}}
\addtocounter{address}{1}
\altaffiltext{\theaddress}{
   Departamento de Astronom\'{\i}a y Astrof\'{\i}sica, Facultad de F\'{\i}sica,
   Pontificia Universidad Cat\'olica de Chile, Casilla~306, Santiago~22, Chile.
\label{CChile}}
\addtocounter{address}{1}
\altaffiltext{\theaddress}{Fermi National Accelerator Laboratory, P.O. Box 500,
   Batavia, IL 60510.
\label{FNAL}}
\addtocounter{address}{1}
\altaffiltext{\theaddress}{Astronomy and Astrophysics Center, University of
   Chicago, 5640 South Ellis Avenue, Chicago, IL 60637.
\label{Chicago}}
\addtocounter{address}{1}
\altaffiltext{\theaddress}{Adler Planetarium, Chicago, IL 60605.
\label{Adler}}
\addtocounter{address}{1}
\altaffiltext{\theaddress}{Department of Physics and Astronomy, University
   of Pittsburgh, 3941 O'Hara Street, Pittsburgh, PA 15260.
\label{Pittsburgh}}
\addtocounter{address}{1}
\altaffiltext{\theaddress}{Department of Astronomy, University of Washington,
   Box 351580, Seattle, WA 98195.
\label{UW}}
\addtocounter{address}{1}
\altaffiltext{\theaddress}{Microsoft Research, 455 Market Street, Suite 1690,
   San Francisco, CA 94105.
\label{MicroSoft}}
\addtocounter{address}{1}
\altaffiltext{\theaddress}{Max-Planck-Institute f\"ur extraterrestrische
   Physik, Geissenbachstrasse.~1, D-85741 Garching, Germany.
\label{MPE}}
\addtocounter{address}{1}
\altaffiltext{\theaddress}{Department of Physics, New York University,
   4 Washington Place, New York, NY 10003.
\label{NYU}}
\addtocounter{address}{1}
\altaffiltext{\theaddress}{Apache Point Observatory, P.O. Box 59,
   Sunspot, NM 88349-0059.
\label{APO}}
\addtocounter{address}{1}
\altaffiltext{\theaddress}{Astronomy Department, California Institute of
   Technology, Pasadena, CA 91125.
\label{Caltech}}
\addtocounter{address}{1}
\altaffiltext{\theaddress}{Department of Physics, Massachusetts Institute
   of Technology, Cambridge, MA 02139.
\label{MIT}}
\addtocounter{address}{1}
\altaffiltext{\theaddress}{Institut d'Estudis Espacials de Catalunya/CSIC,
Gran Capita 2-4, 08034 Barcelona, Spain.
\label{Barcelona}}
\addtocounter{address}{1}
\altaffiltext{\theaddress}{Department of Astronomy and Research Center for the
   Early Universe, School of Science, University of Tokyo, Mitaka,
   Tokyo 181-0015, Japan.
\label{Tokyo2}}
\addtocounter{address}{1}
\altaffiltext{\theaddress}{Institute for Cosmic Ray Research, University
   of Tokyo, Kashiwa, 2778582, Japan.
\label{CosJapan}}
\addtocounter{address}{1}
\altaffiltext{\theaddress}{The Institute for Advanced Study, Princeton,
   NJ 08540.
\label{IAS}}
\addtocounter{address}{1}
\altaffiltext{\theaddress}{Department of Physics and Astronomy,
   The Johns Hopkins University, 3400 North Charles Street,
   Baltimore, MD 21218.
\label{JHU}}
\addtocounter{address}{1}
\altaffiltext{\theaddress}{US Naval Observatory, 3450 Massachusetts Avenue NW,
   Washington, DC 20392-5420.
\label{USNODC}}
\addtocounter{address}{1}
\altaffiltext{\theaddress}{H.N. Russell Fellow.
\label{HNRus}}
\addtocounter{address}{1}
\altaffiltext{\theaddress}{Enrico Fermi Institute, The University of Chicago,
  5640 South Ellis Avenue, Chicago, IL 60637.
\label{Fermi}}
\addtocounter{address}{1}
\altaffiltext{\theaddress}{Lawrence Berkeley National Laboratory,
One Cyclotron Rd., Berkeley CA 94720-8160.
\label{LBNL}}
\addtocounter{address}{1}
\altaffiltext{\theaddress}{Astronomy Centre, University of Sussex, Falmer,
   Brighton BN1~9QJ, UK.
\label{Sussex}}
\addtocounter{address}{1}
\altaffiltext{\theaddress}{Space Telescope Science Institute,
   3700 San Martin Drive, Baltimore, MD 21218.
\label{STScI}}
\addtocounter{address}{1}
\altaffiltext{\theaddress}{Royal Observatory, Edinburgh, Blackford Hill,
   Edinburgh EH9~3HJ, UK.
\label{Edinburgh}}
\addtocounter{address}{1}
\altaffiltext{\theaddress}{US Naval Observatory, Flagstaff Station,
   P.O. Box 1149, Flagstaff, AZ 86002-1149.
\label{USNOAZ}}
\addtocounter{address}{1}
\altaffiltext{\theaddress}{Department of Physics, Applied Physics and
   Astronomy, Rensselaer Polytechnic Institute, Troy, NY 12180.
\label{RPI}}
\addtocounter{address}{1}
\altaffiltext{\theaddress}{Department of Physics, Carnegie Mellon University,
     5000~Forbes Ave., Pittsburgh, PA~15232.
\label{CMU}}
\addtocounter{address}{1}
\altaffiltext{\theaddress}{Physics Department, Rochester Institute of
 Technology, 85 Lomb Memorial Drive, Rochester, NY 14623-5603.
\label{RIT}}
\vbox{
\begin{abstract}
We present the second edition of the Sloan Digital Sky Survey (SDSS)
Quasar Catalog.  The catalog consists of the~16,713 objects
in the SDSS First Data Release that have luminosities larger
than \hbox{$M_{i} = -22$} (in a cosmology with
\hbox{$H_0$ = 70 km s$^{-1}$ Mpc$^{-1}$,}
\hbox{$\Omega_M$ = 0.3,}
and \hbox{$\Omega_{\Lambda}$ = 0.7),} have at least one
emission line with FWHM larger than 1000~km~s$^{-1}$, and have highly
reliable redshifts.
The area covered by the catalog is~$\approx$~1360~deg$^2$.
The quasar redshifts range from~0.08 to~5.41, with a median value of~1.43.
For each object the catalog presents positions accurate
to better than~0.2$''$~rms per coordinate,
five-band ($ugriz$) CCD-based photometry with typical accuracy
of~0.03~mag, and information on the morphology and selection method.
The catalog also contains some radio, near-infrared, and X-ray emission
properties of the quasars, when available, from other large area surveys.
Calibrated digital spectra
of all objects in the catalog, covering the wavelength
region
\hbox{3800--9200 \AA\ } at
a spectral resolution \hbox{of 1800--2100,} are available.
This publication supersedes the first SDSS Quasar Catalog, which was
based on material from the SDSS Early Data Release.  A summary of corrections
to current quasar databases is also provided.

The majority of the objects were found
in SDSS commissioning data using a multicolor selection technique.
Since the quasar selection algorithm was undergoing testing during the
entire observational period covered by this catalog,
care must be taken when assembling samples
from the catalog for use in statistical studies.
A total of~15,786 objects (94\%) in the catalog were discovered by the SDSS;
12,173 of the SDSS discoveries are reported here for the first time.  Included
in the new discoveries are five quasars brighter than \hbox{$i = 16.0$} and
17 quasars with redshifts larger than~4.5.

\end{abstract}
}


\section{Introduction}


Over the past two decades the number of quasars produced by an individual
survey has increased
by about two orders of magnitude.  One of the most intensively studied
quasar data sets, the Bright Quasar Survey of Schmidt \& Green~(1983), contains
approximately 100 quasars. In the 1990s the Large Bright Quasar Survey
(LBQS; Hewett, Foltz, \& Chaffee 1995, 2001) presented more than 1,000
quasars, and recent results from  the 2dF Quasar Survey (Croom et al.~2001)
have pushed quasar survey sample sizes past the 10,000 object milestone.

This paper describes the second edition of the Sloan Digital Sky Survey
(SDSS) Quasar Catalog.  The goal of the SDSS quasar survey is to
obtain spectra of~$\approx$~100,000 quasars from 10,000~deg$^2$ of
high Galactic latitude sky.
The survey will provide CCD-based photometry
in five broad bands covering the entire optical window,
morphological information, and spectra from
\hbox{3800--9200 \AA\ at} a spectral resolution \hbox{of 1800--2100.}
A review of the SDSS is given by York et al.~(2000); Richards et al.~(2002)
present an overview of the quasar target selection
algorithm, and the first edition
of the SDSS Quasar Catalog, based on the SDSS Early Data Release
(EDR; Stoughton et al.~2002),
is given by Schneider et al.~(2002, hereafter Paper~I).

The catalog in the present paper
consists of the 16,713 objects in the SDSS First Data Release
(DR1; Abazajian et al.~2003) that
have a luminosity larger than
\hbox{$M_{i} = -22.0$}  (calculated assuming an
\hbox{$H_0$ = 70 km s$^{-1}$ Mpc$^{-1}$,} \hbox{$\Omega_M$ = 0.3,} 
\hbox{$\Omega_{\Lambda}$ = 0.7} cosmology,
which will be used throughout this paper), and whose SDSS
spectra contain at least one broad emission line
(velocity FWHM larger than \hbox{$\approx$ 1000 km s$^{-1}$).}
The quasars range in redshift from~0.08 to~5.41, and
15,786~(94\%) were discovered by the SDSS.  An object is classified as
previously known if 
the NASA/IPAC Extragalactic Database (NED) Quasar Catalog
contains a quasar within~5$''$ of the SDSS position.
Occasionally NED lists the SDSS designation for an object that was discovered
via another investigation.  In the DR1 catalog we have not attempted
to correct these misattributions.

The DR1 catalog does not include classes of Active Galactic Nuclei (AGN)
such as Type~II quasars and BL~Lacertae objects; studies of these sources
in the SDSS
can be found in Zakamska et al.~(2003) and Anderson et al.~(2003),
respectively.

The objects are denoted in the catalog by their DR1
J2000 coordinates;
the format is \hbox{SDSS Jhhmmss.ss+ddmmss.s}.  Since continual improvements
are being made to the SDSS data processing software, the astrometric
solutions to a given set of observations can result in modifications to the
coordinates of an object at the~0.1$''$ to~0.2$''$ level, hence
the designation of a given source can change between data releases.  Except on
rare occasions (see \S 5.1), this change in position is much less
than~1$''$.
When merging SDSS Quasar Catalogs with previous databases one should
always use the coordinates, not object names, to identify unique entries.
This paper provides a list of corrections to current public quasar databases
(\S 5.13).

The catalog also contains the information required to retrieve the
digital spectra of all the quasars (see \S 4).

The observations used to produce the catalog are presented in
\S 2; the construction of the catalog and the catalog format
are discussed in \S\S 3 and~4, respectively.  Section~5
contains an overview of the catalog, and a brief discussion of future work is
given in \S 6.
The catalog can be found at a public web
site.\footnote{\tt 
http://www.sdss.org/dr1/products/value$\_$added/qsocat$\_$dr1.html}

\section{Observations}

\subsection{Sloan Digital Sky Survey}

The Sloan Digital Sky Survey
uses a CCD camera \hbox{(Gunn et al. 1998)} on a
dedicated 2.5-m telescope
at Apache Point Observatory,
New Mexico, to obtain images in five broad optical bands over
10,000~deg$^2$ of the high Galactic latitude sky, most of which is a
contiguous area centered approximately
on the North Galactic Pole.  The five filters 
(designated $u$, $g$, $r$, $i$, and~$z$) 
cover the entire wavelength range of the CCD
response \hbox{(Fukugita et al.~1996);} updated filter response
curves are given by Stoughton et al.~(2002).

The
survey data processing software measures the properties of each detected object
in the imaging data in all five bands, and determines and applies both
astrometric and photometric
calibrations (Pier et al., 2003; \hbox{Lupton et al. 2001)}.
Version~5.3 of the photometric pipeline software was used
to process the imaging data used for the quasar catalog.
The DR1 images have a median seeing of about~1.5$''$; only 10\% of the
data have seeing larger than~1.7$''$ (the precise values of these numbers
are filter-dependent).
The~50\% completeness limits for point sources in DR1 are
typically~22.5, 23.2, 22.6, 21.9,
and~20.8 in~$u$, $g$, $r$, $i$ and~$z$,
respectively (Abazajian et al.~2003).
The image of an unresolved source brighter
than \hbox{$r \approx 14$} will be saturated.
All magnitudes in the quasar catalog refer to the point spread function
measurements of the SDSS photometric pipeline (see Stoughton et al.~2002 and
Abazajian et al.~2003 for details).

Photometric calibration is provided by simultaneous
observations with a 20-inch telescope at the same site (see Hogg
et al.~2001, Smith et al.~2002, and Stoughton et al.~2002).
The SDSS photometric system is based on the AB magnitude scale
(Oke \& Gunn~1983); the SDSS system was designed so that the colors of
an object with a constant $f_{\nu}$ spectral energy distribution are zero.
We estimate, from several lines of argument, that the
$(g-r)$, $(r-i)$, and $(i-z)$ colors are AB to within~$\approx$~3\%.
The $(u-g)$ color appears to be too red by about~5\%
(in the sense that \hbox{$(u-g)_{\rm AB} \ = \ (u-g)_{\rm SDSS} - 0.05$).}
The photometric measurements are placed on the natural system of the 2.5-m
telescope, and the asterisks affixed to the filter names of the
photometric measurements in the previous catalog (e.g., $u^*$), which
indicated the preliminary status of the calibration of the SDSS system,
have been removed for the DR1 measurements.
The accuracy of the absolute calibration of the magnitudes
(i.e., in Janskys) is now expected to be better
than~10\% (Abazajian et al.~2003).

The catalog contains photometry from~52 different
SDSS imaging runs taken between 
1998~September~19 (Run~94) and 2002~September~5 (Run~3325)
and spectra from 291 spectroscopic plates taken between 
2000~March~5 and 2001~October~21.

\subsection{Target Selection}

The SDSS filter system was designed to identify quasars at redshifts between
zero and six (see
Paper~I and Richards et al.~2002 for an overview of SDSS quasar selection).
Most quasar candidates are selected based on
their location in multidimensional SDSS color-space.
Objects with colors that place them outside of the stellar locus
and do not inhabit some specific ``exclusion" regions (e.g., places
dominated by white dwarfs, A stars, and M star-white dwarf pairs)
are identified as primary quasar candidates.  An $i$ magnitude limit of~19.1
is imposed for
candidates whose colors indicate
a probable redshift of less than~$\approx$~3 (the objects
identified in the $ugri$ color cube, most are ``UV excess" objects at
redshifts of two or less);
high-redshift candidates
(found in the $griz$ color cube) are accepted if \hbox{$i < 20.2$.}
The vast majority (over 90\%) of SDSS-selected quasars
follow a remarkably tight color-redshift relation in the SDSS color-system
(Richards et al.~2001).  
In addition to the multicolor selection, unresolved objects brighter
\hbox{than $i = 19.1$} that lie within~1.5$''$ of a FIRST radio source
(Becker, White, \&~Helfand~1995) are also identified as primary quasar
candidates.

The Point Spread Function (PSF) magnitudes are used for the quasar
target selection, and the selection is based on magnitudes and colors
that have been corrected for Galactic absorption
(using the maps of Schlegel, Finkbeiner, \& Davis~1998).
For candidates whose likely redshifts are less than three,
both extended and point sources are included as quasar candidates
for the spectroscopic program; however,
extended sources are excluded if they lie in a region of color space that is
densely occupied by normal galaxies (see Richards et al.~2002).  At larger
likely redshifts, an object must be unresolved in the SDSS images
to become a multicolor spectroscopic target (in order to reduce contamination
from faint red galaxies).

The PSF magnitudes will suffer a bias for resolved objects; if the luminosity
of resolved component (the host galaxy) is comparable to that of
the point source (the AGN), the PSF magnitude will overestimate the
AGN brightness.
To estimate the
size of this effect, we have examined the difference between the
PSF and ``model" (i.e., total) magnitudes for a set of normal galaxies
with magnitudes between~18 and~21.5.  The PSF magnitude in these cases
represent the
light from the galaxy that would be subsumed in the AGN profile; as one
expects, as the brightness of the galaxy decreases, the difference between
the galaxy
PSF and model magnitudes decline (fainter galaxies tend to be smaller).
An extreme case is an AGN component at the $i = 19.1$ flux limit
and the \hbox{$M_i = -22.0$} luminosity limit
residing in a galaxy whose brightness is equal to the AGN, the reported
PSF magnitude will be~$\approx$~0.40~mag brighter than
that due to the AGN; if the
galaxy's total brightness is two magnitudes below that of the AGN component,
the PSF magnitude for an $i = 19.1$ AGN will be brightened
by~$\approx$~0.09~mag.
This exercise shows that the presence of a significant host galaxies can
cause an overestimate in the AGN numbers in a flux-limited sample,
but host galaxy effects can also remove legitimate AGNs from the
sample; contamination from the galaxy may cause the object to
shift out of the quasar selection region in the color-color diagram.
The calibration of how host galaxies impact the interpretation of the
SDSS quasar sample is clearly a complex
task and beyond the scope of this paper.  (The SDSS
quasar survey is not unique in suffering from this selection effect.)

Supplementing the primary quasar sample described above are
quasars that were targeted by
the following SDSS software selection packages:
Galaxy (part of the SDSS galaxy sample; Strauss et al.~2002 and
Eisenstein et al.~2001),
X-ray (object near the position of a ROSAT All-Sky Survey (RASS)
source; Anderson et al.~2003),
Star (point source with unusual color), or Serendipity (unusual color
or FIRST matches).
No attempt at completeness was made for the last three categories;
objects selected by these algorithms are observed if a given spectroscopic
plate has fibers remaining after all of the high-priority classes (galaxies,
quasars, and sky and spectrophotometric calibrations) in
the field have been assigned fibers
(see Blanton et al.~2003).  Most of the DR1 quasars that
fall below the magnitude limits of the quasar survey were selected by
the serendipity algorithm (see \S 5).

Target selection also imposes a maximum brightness limit on the objects.
Accurate photometric measurements
of point sources brighter than \hbox{$r \approx 14$}
are impossible as their images are saturated; objects that have saturated
pixels are dropped from further consideration.
An additional constraint is introduced to prevent
saturation and fiber cross-talk problems in the SDSS spectroscopic
observations; an object
cannot be included in the quasar spectroscopic program if
it has an~$i$  magnitude
brighter than~15.0.  Objects may also be dropped
from consideration if the photometric measurements are considered
suspect (approximately 5\% of the area);
for example, objects close to very bright stars, data badly
affected by bad CCD columns and CCD saturation trails, etc.

One of the most important tasks during the SDSS commissioning period
was to refine the quasar target selection algorithm (see Paper I); all of
the DR1 data were taken during this period.  The
quasars in the catalog were selected using eleven different versions
of selection software; none of the DR1 quasars were selected for spectroscopy
using the
final version of the code described by Richards et al.~(2002).
Once the final target selection software was
installed, the algorithm was applied to the entire DR1 photometric database.
The DR1 quasar
catalog contains two sets of spectroscopic targeting information for
each object: that used for the spectroscopic observation, and the values
produced by the final algorithm (see \S 4).  It is important to note
that extreme care must be exercised when constructing statistical samples
from this catalog; not only must one drop objects that were not
identified by the final selection software, one must also account for
quasar candidates produced by the final version that were not slated for
spectroscopic observation by earlier versions of the software.
This caveat aside, the selection for the UV-excess quasars,
which comprise the vast majority ($\approx$ 80\%) of the objects in the
DR1 Catalog, has remained
reasonably uniform; the changes to the selection algorithm were primarily
aimed at improving the effectiveness of the identification of \hbox{$z > 2.5$}
quasars.  A more extensive discussion of this issue can be found
in \S 5.2.

The efficiency (the ratio of true quasars to quasar candidates)
and completeness of the final selection algorithm have not
yet been precisely calibrated; both quantities have strong brightness and
redshift dependences.  Preliminary estimates indicate that the design
goals (65\% efficiency and~90\% completeness) should be achieved, although
the use of different selection algorithms causes these values to vary
in the DR1 database.

\subsection{Spectroscopy}

Spectroscopic targets chosen by the various SDSS selection algorithms
(i.e., quasars, galaxies, stars, serendipity) are arranged onto
a series of 3$^{\circ}$ diameter circular fields (Blanton et al.~2003).
A typical spectroscopic field contains about~75 quasar candidate spectra
(the number of confirmed quasars per field in the DR1 catalog ranges from~17
to~162;
commissioning tests are responsible for this large range).
Details of the spectroscopic observations can be found in
York et al.~(2000), Castander et al.~(2001), Stoughton et al.~(2002),
and Paper~I.
The two double-spectrographs produce data covering \hbox{3800--9200 \AA }
and have a spectral resolution ranging from 1800 to 2100.

There are~291 DR1 spectroscopic fields.
The majority of the coverage
lies near the Celestial Equator; the locations of the plate centers
can be found from the information given by Abazajian et al.~(2003).
The total area enclosed by the spectroscopic fields
is 1565~deg$^2$ (there is considerable overlap
between many of the fields).  Not all of the plates, however,
were fully covered by spectroscopic targets, as in some cases SDSS imaging
did not cover the entire plate area at the time of plate drilling;
the actual area with DR1 spectra
is~$\approx$~1360~deg$^2$.

The data, along with the associated calibration frames, are processed by
the SDSS spectroscopic pipeline, which removes
instrumental effects, extracts the spectra, determines the wavelength
calibration, subtracts the sky spectrum, removes the atmospheric
absorption bands, merges the blue and red spectra,
and performs the flux calibration.  
The calibrated spectra are classified into various groups
(e.g., star, galaxy, quasar) and redshifts determined by two independent
software packages (see Stoughton et al.~2002 for a description of the
spectroscopic processing code).  Objects whose spectra cannot be classified
by the software are flagged for visual inspection.
The quasar classification is based
solely on the presence of broad emission lines in the spectra;
the classification software does not
employ information about the selection of the object
(e.g., was the spectrum obtained because the target selection process
identified the object as a quasar candidate?), nor is luminosity used by the
SDSS pipeline
as a criterion for designating an object as a quasar.
The quasar spectra presented herein (and the sample from
which they were drawn) were visually inspected by several of the authors
(DPS, PBH, GTR, MAS, DVB, and SFA)  to 
ascertain the accuracy of the redshifts (see \S 3.2).


Figure~1 shows the calibrated SDSS spectra of six of the catalog quasars
representing a wide range of properties;
all were previously unknown.
These spectra have been rebinned for display purposes.  A composite
quasar spectrum formed from SDSS data is discussed by
Vanden~Berk et al.~(2001).

\section{Construction of the SDSS Quasar Catalog}

The quasar catalog was constructed in three stages: 1)~Creation of a
quasar candidate database, 2)~Visual examination of the candidates'
spectra, and 3)~Application of luminosity
and emission-line velocity width criteria.

The continuous evolution of the photometric pipeline during the time period
when the DR1 data were being acquired creates a situation where the
most accurate photometric measurements may not be the values used for
the actual target selection, i.e., the measured properties of an
object (astrometry, photometry) can change between the time the spectroscopic
plates are drilled and the latest processing of the data.  There are two
sets of measurements that are relevant: 1)~TARGET refers to the values used
at the times that spectroscopic targets were selected for the drilling of
plates (the TARGET selection flags are those produced by the selection
algorithm at the time of plate drilling), and
2)~BEST are the results of the latest processing of the photometric
pipeline (i.e., the BEST selection flag is that produced by the
Richards et al.~2002 algorithm). 
Unless the type of measurement is explicitly given, the BEST
values are reported.

Occasionally the actual image data changes between TARGET and BEST; this
occurs when a previously scanned (and targeted)
region is reobserved under better conditions.
In such a situation the TARGET parameters do not change,
but the BEST values refer to the quantities derived from the
highest quality images.
In this case quasar variability can produce significant changes between
the TARGET and BEST photometry, and introduce changes between the
Paper~I and DR1 photometry.

\subsection{Creation of Quasar Candidate Database}

A quasar candidate database was constructed by requesting the image and
spectroscopic information on all objects with DR1 spectra that were
either targeted (using the TARGET flags)
as a quasar candidate or whose spectra were classified
as quasars by the SDSS software.
The query to the SDSS database
returned a total of~35,535 objects.
This quasar database was constructed prior to the 
final development of the project's Catalog Archive Server~(CAS),
therefore it is 
not possible to recreate this database with straightforward queries 
to the CAS.  The differences are not expected to be significant, but 
future work will use the CAS protocol.

At this stage two additional cuts
were made: 1)~objects whose spectra were
classified with high confidence as ``stars" and had redshifts less
than~0.002 were rejected, and 2)~multiple observations of the same object
(coordinates agreed to 1$''$) were resolved; the spectrum with the highest
S/N ratio was retained.  These culls resulted in the removal of~8345 entries.

This database was supplemented by a visual examination of all
``unclassifiable" DR1 spectra and by investigating objects that appeared
to be quasars but whose SDSS archive entries were for various reasons
incomplete.  These efforts identified twelve quasars that were added
to the quasar database.

\subsection{Visual Examination of the Spectra}

The SDSS spectrum of each of the~27,202 remaining quasar
candidates was manually inspected by
several of the authors.  This effort confirmed that
spectroscopic pipeline redshifts of the vast majority of the objects
were accurate.
Several thousand objects were dropped from the
list because they were obviously not quasars (these objects tended to be
galaxies that had been targeted as a possible quasar by
the selection software).
Spectra for which redshifts could not be determined (low signal-to-noise
ratio or subject to data processing difficulties) were also removed from
the sample, as were a handful of objects with unreliable photometry
(i.e., the photometric flags indicated severe problems with the
measurements).
The visual inspection also resulted in the revisions of
the redshifts of approximately 100 quasars; usually this adjustment was
quite substantial (pipeline misidentification of emission line).

\subsection{Luminosity and Line Width Criteria}

Quasars have historically been defined as the high-luminosity branch of
Active Galactic Nuclei (AGN); the (somewhat arbitrary) luminosity division
between quasars and Seyfert galaxies has a consensus value
of \hbox{$M_B = -23$} for an \hbox{$H_0$ = 50 km s$^{-1}$ Mpc$^{-1}$,}
\hbox{$\Omega_M$ = 1,}
\hbox{$\Omega_{\Lambda} = 0$}
cosmology (e.g., Schmidt and Green 1983).

The luminosity calculations in Paper~I employed this standard cosmology.
Paper~I selected the~$i$ band as the luminosity indicator (i.e.,
required that \hbox{$M_i < -23$})
as this choice offered a
number of advantages over the $B$ filter (or the similar $g$ photometry of
the SDSS):
1)~Galactic absorption is less important in $i$ than $B$,
2)~the ability to detect ``red" or ``reddened" quasars is enhanced
in the $i$ band, and 3)~sensitivity to high-redshift quasars
(at redshifts above about~2.5 the
Lyman~$\alpha$ forest enters the $g$ bandpass, whereas this does not occur
in the $i$ filter until redshifts near five).  For these reasons
the flux limit for the SDSS Quasar Survey is set by the $i$ photometry.
The only significant
disadvantage of this approach is that, relative to $B$ photometry, the
$i$ band measurements contain a larger contribution of host galaxy starlight;
this can be important at low redshift.  This effect
is partially mitigated, however, by the use of
point spread function photometry in the quasar selection, even for
extended sources.
The luminosity division of Paper~I, \hbox{$M_i = -23$,}
corresponds to \hbox{$M_B \approx  -22.65$} for a ``typical" quasar
spectral energy distribution.

The DR1 catalog adopts a cosmology consistent
with the recent results from the
Wilkinson Microwave Anisotropy Probe (WMAP; Bennett et al.~2003):
\hbox{$H_0$ = 70 km s$^{-1}$ Mpc$^{-1}$,} $\Omega_M = 0.3$,
and \hbox{$\Omega_{\Lambda} = 0.7$} (Spergel et al.~2003).
Figure~2 displays the luminosity differences introduced by this change
of cosmological parameters
(using the formulae presented by Hogg~1999).
The change in
$H_0$ produces a lowering of the calculated
luminosity by 0.73~mag at zero redshift.
The luminosities of the WMAP and standard cosmologies
become equal at a redshift of approximately 1.7; at larger redshifts
the WMAP cosmology luminosities become steadily (albeit slowly) larger
with increasing redshift.
The luminosities of high-redshift quasars calculated by the``standard" and
WMAP cosmologies are similar 
because at \hbox{$z > 3$} the effect of the different values of
$\Omega_M,$ and $\Omega_{\Lambda}$ is almost cancelled
by the change of $H_0$ from~50 \hbox{to 70 km s$^{-1}$ Mpc$^{-1}$}
(see Figure~2).

The absolute magnitudes were calculated by correcting the $i$ 
measurement for Galactic extinction (using the maps of
Schlegel, Finkbeiner, \& Davis~1998) and assuming that the quasar
spectral energy distribution in the ultraviolet-optical
can be represented by a power law
\hbox{($f_{\nu} \propto \nu^{\alpha}$),} where $\alpha$~=~$-0.5$
(Vanden~Berk et al.~2001).  This calculation ignores the contributions
of emission lines, and will break down for quasars with redshifts larger
than~5.0 as the strong Lyman~$\alpha$ emission line (typical observed
equivalent width of~$\approx$~400~\AA\ at \hbox{$z \approx 5$;} Fan et al.~2001)
and the Lyman~$\alpha$ forest
enter the~$i$ bandpass.

We have adopted a luminosity cutoff for the DR1 catalog
of $M_i < -22$; this corresponds
to a Paper~I (``standard" quasar cosmology) absolute magnitude of
\hbox{$M_B \approx -22.4$} for an object at zero redshift
with a typical AGN spectrum.  An object
of $M_i = -22$ will reach the ``low-redshift" ($ugri$ selected) SDSS quasar
selection limit at a redshift of~$\approx$~0.4.

For each emission line, the spectroscopic pipeline
returns values for line widths (the FWHM of a single Gaussian).
Quasar emission line profiles are frequently poor matches to a Gaussian,
and the line fitting process fails
in several percent of the cases; when this occurs zero is entered for the
line width.
We have examined the line
profiles of all objects satisfying our luminosity criterion that had
quoted maximum line FWHMs of less than~1000~km~s$^{-1}$, and
included in the catalog
objects that had at least one line (or significant component of a line)
whose ``manual" Gaussian-fit FWHM exceeded~1000~km~s$^{-1}$.  The line
fitting software is currently being modified, and it is hoped that future
editions of the SDSS Quasar Catalog will include reliable, automated
measurements of the maximum emission line FWHM for each object.

\section{Catalog Format}

The DR1 SDSS Quasar Catalog is available in three types of files at the public
web site:
1)~a standard ASCII file with fixed-sized columns,
2)~a gzipped compressed version of the ASCII file (reduces the size
by a factor of nearly five), and 3)~a binary FITS table format.
The following description applies to the standard ASCII file.  All files
contain the same number of columns, but the storage of the numbers differs
slightly in the ASCII and FITS formats; the FITS header contains all of the
required documentation.

The standard ASCII catalog (Table~2 of this paper)
contains information on~16,713 quasars in
a 5.4~megabyte file.
The DR1 format is similar to that of the EDR Quasar Catalog (Paper~I), but the
DR1 Catalog is
considerably larger because of the increase of a factor of over
four in number of objects and the expanded information given for
each object (52 columns in DR1 compared to 32 columns in EDR).

The first~58 lines consist of catalog documentation; this is followed
by~16,713 lines containing
information on the quasars.  There are~52 columns in each line; a summary
of the information is given in Table~1 (the documentation in the catalog
is an expansion of Table~1).  At least one space separates all the
column entries, and, except for the first and last columns (SDSS and NED
object names), all entries are reported in either floating point or
integer format.

Notes on the catalog columns:

\noindent
1) The DR1 object designation, given by the format
\hbox{SDSS Jhhmmss.ss+ddmmss.s}; only the final~18
characters (i.e., the \hbox{``SDSS J"} for each entry is dropped)
are given in the catalog.

\noindent
2-3) These columns contain the J2000 coordinates (Right Ascension and
Declination) in decimal degrees.  The positions for the vast majority of
the objects are accurate to~0.1$''$~rms in each coordinate; the largest
expected errors are~0.2$''$ (see Pier et al~2003).

\noindent
4) The quasar redshifts are listed in column 4 of the catalog; see
Paper I and Stoughton et al.~(2002) for a discussion of the redshift
measurements.

\noindent
5) The database search technique used to find the quasar is coded in this
column. If the object was found by the database search for objects
targeted by the quasar selection algorithm or
the spectrum of the object was classified as a quasar by the
SDSS software, this column contains a ``0"; a ``1" indicates the supplemental
sample of twelve objects
(see section~3.1).

\noindent
6-15) These columns contain the BEST DR1 PSF
magnitudes and errors for each object in the five SDSS filters.
The effective wavelengths of the $u$, $g$, $r$, $i$, and $z$ bandpasses
are 3551, 4684, 6166, 7480, and~8932~\AA, respectively (Stoughton et al. 2002).
The magnitudes are reported in the natural system of the SDSS camera.
The SDSS photometric system is normalized so that the $ugriz$ magnitudes are
approximately on the AB system (Oke \& Gunn~1983);
the accuracy of this transformation is approximately~10\% (see discussion
in Abazajian et al.~2003).  The
values refer to magnitudes measured by fitting the point spread
function to the data (see Stoughton et al.~2002).  The magnitudes
have not been corrected for Galactic reddening.
The quantities are asinh magnitudes (Lupton, Gunn, \& Szalay~1999),
which are defined by

$$  m \ = \ -{2.5 \over \ln 10} \ \left[ \ {\rm asinh} \left({f/f_0 \over
2 b} \ \right) \ + \ \ln b \ \right] $$

\noindent
where $f_0$ is the flux density from a zero AB magnitude object
(3631~Jy) and the quantity
$b$ is a softening parameter.
The SDSS has set $b$, which is dimensionless, such that zero flux corresponds
to magnitudes~24.63, 25.11, 24.80, 24.36, and~22.83
in the $u$, $g$, $r$, $i$,
and~$z$ bands, respectively (Stoughton et al.~2002).
For measurements that are more than
2.5~magnitudes brighter than the zero flux values,
the difference between asinh magnitudes and standard magnitudes (Pogson~1856)
is less than~1\%; for the vast majority of the entries in the catalog
the differences between asinh and standard magnitudes are negligible
(the primary exceptions being the $u$ and $g$ magnitudes of
high-redshift quasars).

\smallskip\smallskip
\vbox{\noindent
16) The Galactic absorption in the $u$ band is based on the maps of
Schlegel, Finkbeiner, \& Davis~(1998).  For an $R_V = 3.1$ absorbing medium,
the absorptions in the SDSS bands are

$$ A_u \ = \ 5.155 \ E(B-V) $$
$$ A_g \ = \ 3.793 \ E(B-V) $$
$$ A_r \ = \ 2.751 \ E(B-V) $$
$$ A_i \ = \ 2.086 \ E(B-V) $$
$$ A_z \ = \ 1.479 \ E(B-V) $$
}

There are some differences in the
SDSS filter response curves as originally designed (Fukugita et al.~1996,
Gunn et al. 1998)
and those in operation at the telescope (Stoughton et al.~2002).  These
changes, particularly the 100~\AA\ redward shift of the $u$ response,
cause small but systematic changes in the estimated Galactic absorption,
but these changes are not large for the objects in the catalog.


\noindent
17) If there is a source
in the FIRST catalog
within~1.5$''$ of
the quasar position, this column contains the FIRST
peak flux density (in mJy) at 20~cm encoded as an AB magnitude

$$ AB \ = \ -2.5 \log \left( {f_{\nu} \over 3631 \ {\rm Jy}} \right) $$

\noindent
An entry of ``0.0" indicates no match to a FIRST source; an entry of
``$-1.0$" indicates that the object does not lie in the region covered by
the FIRST survey.

\noindent
18) The S/N of the FIRST measurement.

\noindent
19) Separation between the SDSS and FIRST coordinates (in arc seconds).

\noindent
20) The logarithm
of the vignetting-corrected count rate (photons s$^{-1}$)
in the broad energy band \hbox{(0.1--2.4 keV)} in the
ROSAT All-Sky Survey Faint Source Catalog (Voges et al.~2000) and the
ROSAT All-Sky Survey Bright Source Catalog (Voges et al.~1999).
The matching radius was set to~30$''$;
an entry of~``0.0" in this column indicates no X-ray detection.

\noindent
21) The S/N of the ROSAT measurement.

\noindent
22) Separation between the SDSS and ROSAT All-Sky Survey
coordinates (in arc seconds).

\noindent
23-28) These columns contain the JHK magnitudes and errors from the
2MASS All-Sky Data Release Point Source Catalog (Cutri et al.~2003) using
a matching radius
of~3.0$''$.  A non-detection by 2MASS is indicated by a ``0.0" in these
columns.  Note that the 2MASS measurements are Vega-based, not AB,
magnitudes.

\noindent
29) Separation between the SDSS and 2MASS coordinates (in arc seconds).

\noindent
30) The absolute magnitude in the $i$ band calculated assuming
\hbox{$H_0$ = 70 km s$^{-1}$ Mpc$^{-1}$,}
$\Omega_M$~=~0.3, and $\Omega_{\Lambda}$~=~0.7, a power
law (frequency) continuum index of~$-0.5$, and correcting the $i$ measurement
for Galactic extinction.

\noindent
31) If the SDSS photometric pipeline classified the image of the quasar
as a point source, the catalog entry is~0; if the quasar is extended, the
catalog entry is~1.

\noindent
32) The version of the quasar target selection algorithm used to
select the object is coded in this column  (see the documentation in
the catalog and Stoughton et al.~2002 for details).

\noindent
33-39) These six columns
indicate the spectroscopic target selection status for each object.
This is the information used at the time of the drilling of the spectroscopic
plate.
An entry of~``1" indicates that the object satisfied the given criterion
(see Stoughton et al.~2002 for details).  Note that an object can
be targeted by more than one selection algorithm.   (These entries
are the TARGET selection flags.)

\noindent
40-46) Similar to columns 33-38, but containing the information
for the version of the
Quasar Target Selection Algorithm presented in Richards et al.~2002
(the BEST selection flags).

\noindent
47-48) The SDSS Imaging Run number and the Modified Julian Date of the
photometric observation used in the catalog.  The Modified Julian Date (MJD)
is given as an integer; the catalog entry is the truncated MJD,
and all observations on a given night have the same integer MJD
(and, because of the observatory's location, the same UT date). For example,
imaging run 94 has an MJD of 51075; this observation was taken on the 
night of 1998 September 19 (UT).

\noindent
49-51) Information about the spectroscopic observation (modified Julian
date, spectroscopic plate number, and spectroscopic fiber number) used to
determine the redshift are contained
in these columns.  These three numbers are unique for each spectrum, and
can be used to retrieve the digital spectra from the public SDSS database.

\noindent
52) If there is a source in the NED quasar database within~5.0$''$ of the
quasar position, the NED object name is given in this column.
The matching was done using
the 30,477 objects in the NED quasar database as of January~2003.

\section{Catalog Summary}

Of the 16,713 objects in the catalog, 15,786 were discovered by the SDSS and
12,173 are presented here for the first time.
The catalog quasars span a wide range of properties: redshifts
from~0.08 to~5.41, \hbox{$ 15.15 < i < 21.79$}
(84~objects \hbox{have $i > 20.5$;} only three
have \hbox{$i > 21.0$}),
and \hbox{$ -30.3 < M_{i} < -22.0$.}
The catalog contains 1193, 824, and 2260
matches to the FIRST, RASS, and 2MASS catalogs, respectively,
The RASS and 2MASS catalogs cover essentially all of the DR1 area, but 17\% of
the entries in the DR1 catalog lie outside of the FIRST region.

Figure~3 displays the distribution of the DR1 quasars in the $i$-redshift plane
(the three objects with \hbox{$i > 21$} are not plotted).
Objects which NED indicates were previously discovered by investigations other
than the SDSS
are indicated with open circles.  The curved cutoff on the left
hand side of the graph is due to the minimum luminosity criterion
\hbox{($M_i < -22$).}  The ridge in the contours at
\hbox{$i = 19.1$} for redshifts below three reflects the flux limit of the
low-redshift sample; low-redshift points fainter than \hbox{$i = 19.1$}
primarily represent objects selected via criteria other than the primary
multicolor sample (e.g., serendipity).  Above a redshift
of~$\approx$~3, the vast majority of
the catalog quasars were discovered by the~SDSS.

A histogram
of the catalog redshifts is shown in Figure~4.  Most of
the quasars have redshifts below two (the median redshift is~1.43),
but there is a significant tail
of objects out to a redshift of five
\hbox{($z_{\rm max}$ = 5.41).}  The dips in the curve at redshifts
of~2.7 and~3.5 arise because the SDSS colors of quasars at these redshifts
are similar to the colors of stars, so we must accept significant
incompleteness at these redshifts to avoid being flooded by stars masquerading
as quasars.
We expect the final target selection algorithm
(Richards et al.~2002) will reduce the dip seen at
\hbox{$z \approx 3.5$.}

The distribution of the observed $i$ magnitude
(not corrected for Galactic extinction) of the quasars is given in Figure~5.
The sharp drops in the histogram at $i \approx 19.1$ and
$i \approx 20.2$ are due to the magnitude limits in the low and
high redshift samples, respectively.

\subsection{Differences with the SDSS EDR Catalog}

The EDR Catalog (Paper~I) contained 3814 objects.  All of the EDR survey area
is included in the new catalog.  There
are two entries in the EDR catalog that require special comment:

\noindent
{\bf SDSS J004705.83$-$004819.5:} The DR1 processing of this spectrum
revealed that the strong emission line at 4300~\AA\ in the EDR spectrum was
a defect in the data; the continuum is essentially featureless, and this
object is not in the DR1 catalog.  This source should be dropped from
all quasar databases.

\noindent
{\bf SDSS J172543.02+582110.8:} An error was introduced by the EDR
software when matching this spectrum to an object in the image database.
There is an object at this position, but the fiber was drilled using the
information from \hbox{SDSS J172542.16+582110.5} (the designation of the quasar
in the DR1 catalog).  It is not understood
how the~7$''$ offset was introduced in the EDR matching for this object
(all the other
EDR quasars match the DR1 positions within 1$''$).
\hbox{SDSS J172543.02+582110.8} should be dropped from all quasar databases
and be replaced with the DR1 information of {SDSS J172542.16+582110.5.}

The DR1 redshifts of four quasars
(\hbox{SDSS J014905.28$-$011405.0,}
\hbox{SDSS J023044.81$-$004658.0,}
\hbox{SDSS J151307.26$-$000559.3,} and
\hbox{SDSS J171930.24$-$584804.7)}
differ by more than~0.1 from the EDR values; the DR1 measurements are
more reliable.

Thirty of the DR1~$i$
measurements differ by more than~0.5~mag from the corresponding
Paper~I entries.
The source of this discrepancy is almost certainly quasar variability,
as in all~30 cases the photometry presented in the catalog is based
on image data that was taken at a different time than the photometric
observations used in
Paper~I.   For most of the Paper~I quasars, the EDR and DR1 photometry
is based on the same imaging data.  The difference between the EDR and
DR1 $i$ photometry for these objects is roughly Gaussian with a dispersion
of a percent or two; this difference arises from upgrades in the photometric
pipeline.

\subsection{Analysis of Quasar Selection}

The Quasar Selection Algorithm was being tested throughout the period of
DR1 data acquisition; this is reflected in the large number of versions
of the algorithm listed in the catalog.  Although all of the versions
strongly resembled the final selection software (Richards et al.~2002), the
subtle differences in selection demand that caution be exercised (especially
at redshifts above three) when
using the DR1 catalog for statistical studies.

The catalog contains two sets of target selection flags: 1)~the
flag employed to drill the spectroscopic plates (i.e., the information
used to select the object for spectroscopic observation), and 2)~the
target flag produced for each object using the final version of the
selection software.  To create a uniform sample would require taking into 
account those
objects selected by the final target software that were missed by the
code in use at the time of plate drilling.

Therefore the quasars presented in this paper do not represent a uniform and
homogeneous sample, and significant effort will be required to
quantify this to allow it to be used for statistical analysis.  Future
papers will discuss (and resolve) these problems in full; here we
simply describe the issues involved.


In particular, the color
inclusion cuts described in Richards et al.~(2002) are
not implemented for any of the DR1 spectroscopic data.
This means that the quasar target list from the TARGET
imaging data and that from the BEST imaging data
can differ significantly.  In addition,
for some regions of sky 
different imaging runs are used to produce
TARGET and BEST measurements.  For example, the EDR data in the ``Southern
stripe", the images taken on the Celestial Equator in the
Southern Galactic Cap, were among the
earliest observations taken by the SDSS, and they have subsequently been
replaced by data of superior quality.
Photometric quasar variability between the runs
(Vanden~Berk et al.~2003) can cause individual
objects to move in and out of the sample.
Many of the quasars included in this paper (especially
at the faint end) were not selected by the quasar target selection
algorithm at all, but by the Serendipity and RASS match algorithms
(Stoughton et al. 2002; Anderson et al. 2003) and more rarely by
the galaxy target selection algorithms (Strauss et al. 2002;
Eisenstein et al. 2001).


We have examined the nature of those objects targeted as quasar
candidates in the BEST database which do not have spectra.  The
majority of these are missing simply because the DR1 spectroscopic
plates do not fully cover the DR1 imaging area.  The tiling algorithm
that assigns targets to spectroscopic plates (Blanton et al. 2003) is
not fully efficient, causing an additional (small) incompleteness. Of
the remaining objects, we find the DR1 sample is $>85\%$ complete for
bright UV excess sources ($u-g<0.4$), but the completeness of redder objects
is complex and remains to be properly quantified.

The 18 AGN of Schmidt and Green~(1983) that lie within
the DR1 imaging data provide a simple test of the UV excess detection
code.  Three of these objects are brighter than the
$i=15$ spectroscopic survey limit, but the remaining~15 were all identified
by the BEST selection algorithm as multicolor quasar candidates in the primary
survey.  The DR1 Quasar catalog does not contain all~15 of these objects as the
area covered by spectroscopic survey considerably lags (as it must) that of
the imaging survey, and several of the objects fall below our (and that
of Schmidt and Green) quasar luminosity limit.

Because of time-critical demands for spectroscopic plates,
some regions of sky with poor seeing were targeted, with
predictably poor results in quasar target selection.  In addition, we
have a posteriori culled from the quasar target list regions of sky in
which the numbers of quasar candidates appeared unphysically large
for similar reasons of substandard seeing;
these objects were never tiled and therefore never observed
spectroscopically.  We have not yet completed the book-keeping required to
document these various problems. 


In short, a full statistical analysis of the quasar sample will
require analysis of the completeness at the depth described for the
galaxy sample in Appendix~A of Tegmark et al.~(2003).  Our future work
will concentrate on defining the selection function for the DR1 quasar
sample, performing an analysis of those objects selected by the new
algorithm but for which we do not currently have spectroscopy, and
producing a new sample that is suitable for statistical analysis.

A summary of the spectroscopic selection, for both the TARGET and BEST
algorithms, is given in Table~3.  We report
six selection classes, which are columns~33 to~38 (TARGET) and~39 to~44
(BEST) in the catalog.
The second and fourth columns in Table~3 give the number of objects that
satisfied a given selection criterion; the third and fifth columns
contain the number
of objects that were identified only by that selection class.
As expected, the solid majority~(77\%) were selected based on the SDSS
quasar selection criteria;
one-third of the catalog objects were
selected on that basis only.  Approximately~60\% of the quasars
were identified by the serendipity code, which is also primarily an ``unusual
color" algorithm.  About one-fifth of the catalog was selected by
the serendipity criteria alone; these objects tend to be 
$z<3$ quasars that fall below the magnitude limit
of the quasar survey algorithm.

Of the~10,996 DR1 quasars that have Galactic-absorption corrected
$i$ magnitudes brighter than~19.1,
10,719 were found from the quasar multicolor
selection; if one includes multicolor and FIRST selection,
all but~172 of the 10,996 objects are selected.
(If one examines the BEST instead of the
TARGET target flag, 10,310 of the DR1 quasars with Galactic-absorption
corrected $i$ magnitudes brighter than~19.1 were selected
by the multicolor selection.)


\subsection{Bright Quasars}

The catalog contains~19 quasars that have \hbox{$i < 16.0$};
five are reported here for the first time, one is incorrectly
listed by NED as an SDSS discovery (HS1700+6416),
and~13 are non-SDSS discoveries.  The relatively
large number of bright quasars is not produced by the change in
the luminosity cutoff from Paper~I; only one of the 19 objects falls below
the luminosity limit of the Paper~I cosmology.
\hbox{SDSS J073733.01+392037.4} \hbox{($i = 15.98$}, \hbox{$z = 1.74$)} and
\hbox{SDSS J210001.24$-$071136.3} \hbox{($i = 15.15$}, \hbox{$z = 0.60$)} are
two particularly interesting cases; the spectrum of the first object
is displayed in Figure~1.  Of the
193 catalog quasars with \hbox{$i < 17.0$}, 109 are SDSS discoveries. 

Recall that objects that have  \hbox{$i < 15.0$} are not observed with the SDSS
spectrographs; thus 3C~273, which lies in the DR1 area, is not an entry
in the DR1 Quasar Catalog.  Accurate photometric information for known quasars
fainter than the saturation limit \hbox{($i \approx 14$)} in the DR1 area
can be obtained from the imaging observations.

\subsection{Luminous Quasars}

There are 23 catalog quasars with \hbox{$M_{i} < -29$}
(3C~273 has \hbox{$M_{i} \approx -26.6$} in our adopted cosmology); nine were
previously unknown, including
\hbox{SDSS J073733.01+392037.4}, which was mentioned in \S 5.3.
The most luminous quasar in the catalog,
HS~1700+6416 \hbox{(= SDSS J170100.62+641209.0)} at
\hbox{$M_i = -30.25$} and \hbox{$z = 2.73$}, was included in Paper~I.
The most luminous of the previously unknown
quasars have \hbox{$M_i \approx -29.3$.}

\subsection{Broad Absorption Line Quasars}

The SDSS Quasar Selection Algorithm has proven to be effective
at finding Broad Absorption Line (BAL) Quasars.
The spectrum of
\hbox{SDSS J092819.28+534024.2,} a \hbox{$z = 4.39$} quasar with deep,
broad, complex absorption troughs, is displayed in Figure~1.
The EDR BAL catalog, based on
the material in Paper~I, is presented by Reichard et al.~(2003a); 
approximately 6\% of the EDR quasars are BALs.
At redshifts where BALs can be selected from the SDSS, the raw BAL fraction
is~15\% (Tolea et al.~2002) and the fraction after accounting for
color-dependent selection effects is~13.4$\pm$1.2\% (Reichard et al.~2003b).
After including incompleteness corrections, the true fraction of BALs
in the quasar population has been measured as~22$\pm$4\% in the LBQS
(Hewett \& Foltz 2003) and estimated as~15.9$\pm$1.4\% in the SDSS 
(Reichard et al. 2003b).  An investigation of the properties of the Paper~I
BALs is also given in Reichard et al.~(2003b).  Finally, the SDSS
has also demonstrated that it can identify an appreciable
number of ``extreme BALs";
18 of the 23 ususual BALs presented in
Hall et al.~(2002a) are in the DR1 catalog.

\subsection{Quasars with Redshifts Below 0.2}

The catalog contains 101 quasars with redshifts below~0.2; 75 are SDSS
discoveries, and 64 are presented here for the first time.  Approximately
half of these objects have
luminosities below the Paper~I limit.  Figure~1 displays the spectrum of
\hbox{SDSS J091955.33+552137.0,} a \hbox{$z = 0.12$}, \hbox{$M_i = -23.4$}
quasar.  The object with the lowest redshift among the new discoveries is
\hbox{SDSS J142748.28+050222.0} at \hbox{$z = 0.11$}.

\subsection{High-Redshift ($z \ge 4$) Quasars}

There are 186 quasars in the DR1 Quasar Catalog that have
$z \ge 4.0$; 73 are reported here
for the first time.  The SDSS has identified quasars to $z = 6.4$
(e.g., Fan et al.~2003 and references therein), but
objects with $z > 5.7$ cannot be found by the standard SDSS software.
At these redshifts, the observed wavelength of the
Lyman~$\alpha$ emission line is redward of
the $i$ band, and quasars become single-filter ($z$) detections.
At the typical $z$-band flux levels for redshift six quasars, there are simply
too many ``false-positives" to undertake automated targeting.
The largest redshift in the DR1 catalog is
\hbox{SDSS J023137.65$-$072854.5} \hbox{at $z = 5.41$}, which was described by
Anderson et al.~(2001); that paper also describes the properties of
many of the high-redshift objects in the DR1 catalog.
Although only a small fraction of the objects (about 1\%)
in the catalog have redshifts larger than 4.0,
the SDSS is clearly very effective at discovering
such quasars; most of the~186 DR1 catalog entries with \hbox{$z > 4$}
are listed by NED as SDSS discoveries.

The spectra of three interesting new
discoveries are displayed in Figure~1:
1)~\hbox{SDSS J001714.67$-$100055.4} is the only new
$z>5$ quasar reported in this paper;
\hbox{2) SDSS J004054.65$-$091526.8}, at \hbox{$z$ = 4.98,}
is an example
of a weak-lined high-redshift quasar (see e.g., Fan et al.~1999); the
redshift is primarily based on the location of the Lyman~$\alpha$ forest;
and \hbox{3) SDSS J092819.28+534024.2} is a \hbox{$z$ = 4.39} quasar
with strong, complex, broad absorption features.

Although none of the $z >4$ quasars is detected in the RASS, three, including
the previously mentioned weak-lined quasar SDSS~J004054.65$-$091526.8,
serendipitously lie in publicly available fields observed with the
{\it XMM-Newton\/} European Photon Imaging Camera (EPIC).  These three
objects are discussed in the Notes on Individual Objects (\S 5.14).

\subsection{Close Pairs}

The SDSS spectroscopic survey is not particularly effective at identifying
close pairs of quasars because of
the mechanical constraint
that spectroscopic fibers must be separated by~55$''$ on a given plate.
In areas that are covered by more than one plate, however, it is possible
to obtain spectra of both components of a close pair, and
there are~41 pairs of quasars in the catalog with angular separation less than
$60''$ (six pairs with separations less than 20$''$).
Two pairs appear to be physically associated systems ($\Delta z < 0.01$):
\hbox{SDSS J025959.68+004813.6} and
\hbox{SDSS J030000.57+004828.0}, separated by~20$''$ at $z=0.893$;
\hbox{SDSS J085625.63+511137.0} and
\hbox{SDSS J085626.71+511117.8}, separated by~22$''$ at $z=0.543$.
Both of these pairs were included in Paper~I.

\subsection{Morphology}

The SDSS photometric pipeline classifies the images of~821 catalog quasars as
resolved.  As one would expect, the vast majority~(94\%)
of the extended objects
have redshifts below one, but there are a number of resolved quasars at higher
redshifts (the highest redshift of an extended object is~3.58).
A larger fraction of the DR1 quasars are resolved (4.9\%) than in the
EDR sample (3.2\%); this is due to the inclusion of lower
luminosity AGN in the DR1 list (73\% of the quasars with
\hbox{$M_i > -22.2$} are resolved).
The majority of the large redshift ``resolved" quasars are probably measurement
errors, but this sample may contains a few chance superpositions
of quasars and foreground objects or possibly some
small angle separation gravitational lenses.

Two unresolved (in SDSS imaging) quasars have been shown to be
likely gravitational lenses:
\hbox{SDSS J165043.44+425149.3} (Morgan et al.~2003) and
\hbox{SDSS J013435.66-093102.9} (\hbox{PMN J0134$-$0931;}
Gregg et al.~2002, Winn et al.~2002, Hall et al.~2002b).
SDSS~J091301.03+525928.9, a DR1 ``resolved" quasar at a redshift of 1.38,
is the gravitational lens
SBS~0909+532 (Kochanek et al.~1997).
Detailed studies of
three additional DR1 resolved quasars,
\hbox{SDSS J090334.94+502819.3} (Johnston et al.~2003),
\hbox{SDSS J092455.79+021924.9} (Inada et al.~2003a), and
\hbox{SDSS J122608.02$-$000602.2} (Inada et al.~2003b) suggest that they
are excellent candidates for gravitational lenses.

\subsection{Radio Matches}

A radius of~1.5$''$ was used to match DR1 quasars and FIRST
(Becker, White, \& Helfand~1995) sources (this is the same matching radius
used in the SDSS code that selects FIRST sources for spectroscopy;
see Ivezi\'c et al.~2002 for a detailed discussion).
A histogram of the SDSS/FIRST positional offsets of the~1193 matches is
displayed in Figure~6.  The distribution peaks at a matching radius
of~0.2$''$ (similar to the
expected accuracy of the SDSS astrometry), and virtually all of the
matches coincide to better than 1$''$.  Each FIRST identification can
be assigned with extremely high confidence; shifting the declinations of
the DR1 quasar positions by $\pm 200''$ produces zero matches within
a~1.5$''$ radius.  Ivezi\'c et al.~(2002) present
an analysis of the radio properties of SDSS-selected quasars.  The
radio-brightest DR1 quasars that were not previously known have 20~cm
peak flux densities of~$\approx$~700~mJy.

The number of FIRST matches considerably exceeds (by nearly a factor of two)
the number of quasars
selected for spectroscopy by their radio properties (the FIRST target
selection flag).  This occurs for two reasons:~1)~the FIRST flag is set
only for unresolved sources and objects brighter than \hbox{$i = 19.1$}, and
2)~the SDSS spectroscopic targets
for some plates were selected before FIRST information was available for
the region.

\subsection{X-Ray Matches}

Matches with the ROSAT All-Sky Survey Bright (Voges et al. 1999) and
Faint (Voges et al. 2000) Source Catalogs
were made with a maximum allowed positional offset
of~30$''$.  The DR1 catalog lists a total of 824 RASS matches; the positional
offsets are displayed in Figure~7.  The distribution, which peaks
around 10$''$, clearly indicates that the vast majority of identifications
are correct.  Changing the DR1 quasar declinations
by $\pm 200''$ yielded~8 and~14
matches, suggesting that spurious X-ray identifications comprise
at most a few percent of catalog entries.  Vignali et al.~(2003a) and
Anderson et al.~(2003) discuss the X-ray properties of SDSS-selected quasars.

\subsection{Infrared Matches}

The distribution of the angular offsets for the~2260 matches to the
2MASS All-Sky Data Release Point Source Catalog is shown in Figure~8.  The
relatively small fraction (14\%) of matches is because of the
large difference in
survey limits (a typical quasar at the SDSS low-redshift
quasar magnitude limit
will \hbox{be 1--2} magnitudes fainter than the 2MASS detection level).
Virtually all of the identifications are correct; the matching
radius was~3$''$, and~90\% of the SDSS/2MASS coordinates agree to better
than~1$''$ (compare the 2MASS offsets 
with Figure~6, the angular offset distribution for
the FIRST identifications, where the matches are essentially~100\% accurate).

\subsection{Redshift Disagreements with Previous Measurements}

The redshifts of 26 quasars in this catalog disagree by more than~0.10
from the values given in the NED database; the information for each
of these objects is given in Table~4.  The discrepant redshifts of
two of the objects in Table~4
(\hbox{SDSS J002411.65$-$004348.0} and
(\hbox{SDSS J120548.48+005343.8}) were also noted in Paper~I.
In addition, Table~4 also contains the two EDR problems (the non-quasar and
the~7$''$ discrepancy) mentioned in
\S 5.1.  We believe that the DR1 values represent the most accurate
measurements of the redshift.

\subsection{Notes on Individual Objects}

A quick examination of the catalog reveals a few objects that merit
special comment but don't warrant a separate publication.

\noindent
{\bf SDSS~J004054.65$-$091526.8 ($z$ = 4.98):}  The redshift of this
weak-lined quasar
is primarily based on the location of onset of
the Lyman~$\alpha$ forest; the SDSS
spectrum is displayed in Figure~1.
Due to its proximity on the sky to the bright cluster of galaxies Abell~85
($z=0.056$), this object was included in the EPIC
observation of the cluster (see Durret et~al. 2003 for analysis of the 
Abell~85 EPIC data). We have analyzed these data and find a 0.5--2~keV 
source positionally coincident with \hbox{SDSS~J004054.65$-$091526.8};
the probability of 
a false match is only $\approx 3\times 10^{-3}$. Using the EPIC pn data and 
a $50^{\prime\prime}$-radius circular aperture, we measure $31.8\pm 11.6$ net 
counts. The average effective exposure time in the aperture, accounting for 
vignetting, is 3580~s.

Adopting a power-law model with a photon index of 
$\Gamma=2$ and the Galactic column density, we calculate a Galactic 
absorption-corrected flux of $1.3\times 10^{-14}$~erg~cm$^{-2}$~s$^{-1}$ 
in the observed-frame 0.5--2~keV band. The corresponding value of the 
(rest-frame) 2500~\AA\ to 2~keV power-law slope, $\alpha_{\rm ox}$, is $-1.55$ 
(calculated following the methods of Vignali et~al. 2003b). Comparison with 
Figure~6 of Vignali et~al. (2003b) shows that this value of $\alpha_{\rm ox}$ 
is consistent with those found for other $z>4$ SDSS radio-quiet quasars. There 
is no evidence that the X-ray emission from this weak-line quasar is unusual 
or heavily absorbed (see Fan et~al. 1999); if anything,
\hbox{SDSS~J004054.65$-$091526.8}
appears to be mildly X-ray bright given its optical flux. The angular offset 
between \hbox{SDSS~J004054.65$-$091526.8} and the core of Abell~85
\hbox{is $\approx 14.0^{\prime}$}; 
given this offset, we do not expect strong lensing of
\hbox{SDSS~J004054.65$-$091526.8} since 
it lies significantly outside the cluster's Einstein radius. 

\noindent
{\bf SDSS~J083103.01+523533.5 ($z$=4.44):} This object was first noted by
Anderson et al.~(2001), and it is located only~11.4$'$ from the luminous
\hbox{$z = 3.91$} BAL quasar APM~08279+5255 (Irwin et al.~1998).
\hbox{SDSS J083103.01+523533.5} is detected in
an EPIC observation of APM~08279+5255 (see Hasinger, Schartel, \& Komossa 2002
for analysis of the APM~08279+5255 EPIC data). Performing a 0.5--2~keV analysis 
similar to that for \hbox{SDSS~J004054.65$-$091526.8}, we measure 
$92.3\pm 25.1$ net counts, 
a Galactic absorption-corrected
flux of $4.5\times 10^{-15}$~erg~cm$^{-2}$~s$^{-1}$, and
$\alpha_{\rm ox}=-1.74$. 
The observed $\alpha_{\rm ox}$ value is consistent with those in Figure~6 of 
Vignali et~al. (2003b). 

The SDSS Photometric Pipeline identified APM~08279+5255
\hbox{(SDSS~J083141.71+524517.4),} but the object's~$i$ magnitude
of~14.96 exceeds the
maximum brightness allowed for SDSS spectroscopic observations.  The DR1
quasar catalog contains two additional nearby $z>3.9$ objects:
\hbox{SDSS J083212.37+530327.4} and \hbox{SDSS J083324.58+523954.9},
located~18.7$'$ and~16.5$'$, respectively, from APM~08279+5255.
The areal density of luminous, high-redshift quasars near APM~08279+5255,
$\approx$ 10 degree$^{-2}$, is quite remarkable.

\noindent
{\bf SDSS J110213.69+671045.2 ($z=1.96$):} This radio-quiet quasar
possesses relatively narrow
($\sim$1200 km~s$^{-1}$) \ion{C}{3}] emission, very strong narrow
\ion{He}{2} $\lambda$1640 and \ion{O}{3}] $\lambda$1664 emission,
and relatively narrow, self-absorbed \ion{C}{4} emission.  The quasar's spectrum
is similar to those of Type~II quasars (e.g., Stern et al. 2002,
Zakamska et al.~2003),
but with a considerably stronger continuum.  An X-ray observation is
probably required to determine whether this object is a partially obscured
Type~II quasar, or merely a normal quasar with unusually narrow lines.

\noindent
{\bf SDSS~J140146.53+024434.7 ($z$=4.38):} This high-redshift quasar
is detected in an
EPIC
observation of Abell~1835 (see Peterson et~al. 2001 for analysis of the 
Abell~1835 EPIC data). Performing a 0.5--2~keV analysis similar to that for 
\hbox{SDSS~J004054.65$-$091526.8}, we measure 
$57.2\pm 16.9$ net counts, 
a Galactic absorption-corrected flux of 
$7.6\times 10^{-15}$~erg~cm$^{-2}$~s$^{-1}$, and
$\alpha_{\rm ox}=-1.77$. 
The observed $\alpha_{\rm ox}$ value is consistent with those in Figure~6 of 
Vignali et~al. (2003b), so there is no evidence that the intrinsic UV
absorber apparent in the spectrum causes strong X-ray absorption. 
We do not expect strong lensing of \hbox{SDSS J140146.53+024434.7}
by Abell~1835. 

\section{Future Work}

The 16,713 quasars were identified from~$\approx$~15\% of the proposed
SDSS survey area.  The progress of the SDSS Quasar Survey can be seen in
Figure~9, which displays the cumulative number of SDSS quasars
as a function of observing date (for the DR1 catalog); since the closing
date for DR1 observations, the SDSS has identified an
additional~$\approx$~15,000 quasars.  The publication of the
next edition of the SDSS quasar catalog will coincide with
SDSS Second Data Release, currently expected to occur in~2004.

\acknowledgments

We thank the referee, Paul Hewett, for a number of comments
that significantly improved the paper.
This work was supported in part by National Science Foundation grants
AST-9900703~(DPS, GTR), AST-0307582 (DPS), AST-0307384~(XF), and
AST-0071091~(MAS, MNO), and by
NASA LTSA grant NAG5-13035 (WNB, DPS).
XF acknowledges support from an \hbox{Alfred P. Sloan} Fellowship.
PBH acknowledges support from Fundaci\'{o}n Andes.

Funding for the creation and distribution of the SDSS Archive
has been provided by the Alfred P. Sloan Foundation, the
Participating Institutions, the National Aeronautics and Space
Administration, the National Science Foundation, the U.S.
Department of Energy, the Japanese Monbukagakusho, and the
Max Planck Society.
The SDSS Web site \hbox{is {\tt http://www.sdss.org/}.}
The SDSS is managed by the Astrophysical Research Consortium
(ARC) for the Participating Institutions.  The Participating
Institutions are The University of Chicago, Fermilab, the Institute
for Advanced Study, the Japan Participation Group, The Johns
Hopkins University, Los Alamos National Laboratory, the
Max-Planck-Institute for Astronomy (MPIA), the
Max-Planck-Institute for Astrophysics (MPA), New Mexico
State University, University of Pittsburgh, Princeton University,
the United States Naval Observatory, and the University of
Washington.

This research has made use of 1)~the NASA/IPAC Extragalactic Database (NED)
which is operated by the Jet Propulsion Laboratory, California Institute
of Technology, under contract with the National Aeronautics and Space
Administration, 2)~data products from the Two Micron All Sky 
Survey, which is a joint project of the University of
Massachusetts and the Infrared Processing and Analysis Center/California 
Institute of Technology, funded by the National Aeronautics
and Space Administration and the National Science Foundation, and
3)~on observations obtained with {\it XMM-Newton\/},
an ESA science 
mission with instruments and contributions directly funded by 
ESA Member States and the USA~(NASA).

\newpage





\begin{figure}
\plotfiddle{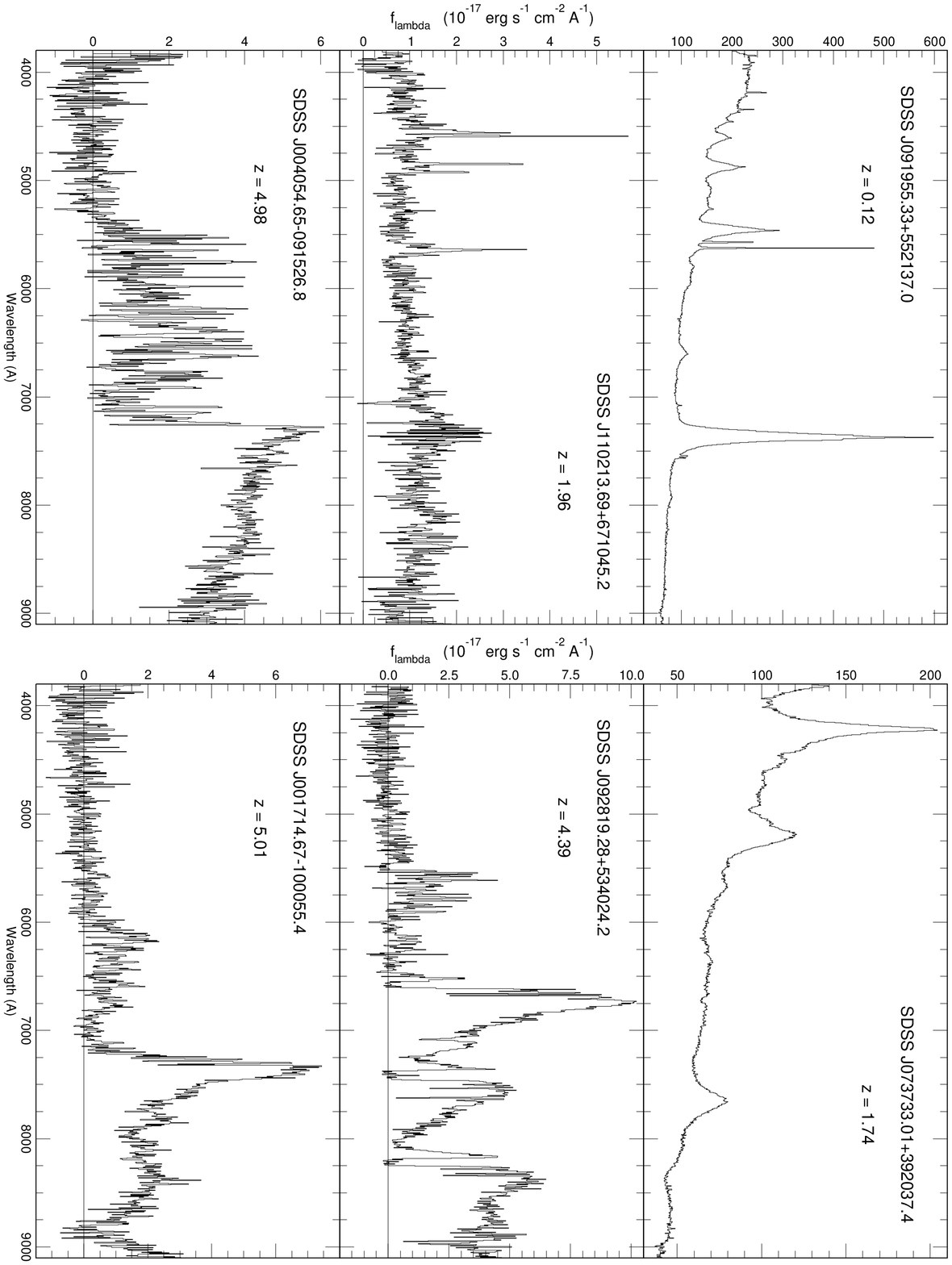}{6.0in}{90.0}{70.0}{70.0}{270.0}{0.0}
\figcaption{
An example of data produced by the SDSS spectrographs.  The spectral
resolution of the data ranges from 1800 to 2100; a dichroic splits the beam
at~6150~\AA .  The data have been rebinned \hbox{to 5 \AA\ pixel$^{-1}$}
for display purposes.  All six of the quasars were discovered by the~SDSS
and reported here for the first time.
Notes on spectra: 1)~\hbox{SDSS J091955.33+552137.0} is a luminous
($M_i = -23.4$), low-redshift quasar;
2)~\hbox{SDSS J073733.01+392037.4} is both bright ($i = 15.98$)
and very luminous \hbox{($M_i = -29.3$)};
3)~\hbox{SDSS J110213.69+671045.2} is a narrow-lined radio quiet quasar;
4)~\hbox{SDSS J092819.28+534024.2} is an example of a BAL with complex
absorption troughs;
5)~\hbox{SDSS J004054.65$-$091526.8} is a weak-lined, high-redshift
quasar; and
6)~\hbox{SDSS J001714.67$-$100055.4} has the highest redshift ($z = 5.01$)
of the newly discovered quasars.
\label{Figure 1 }
}
\end{figure}

\clearpage

\begin{figure}
\plotfiddle{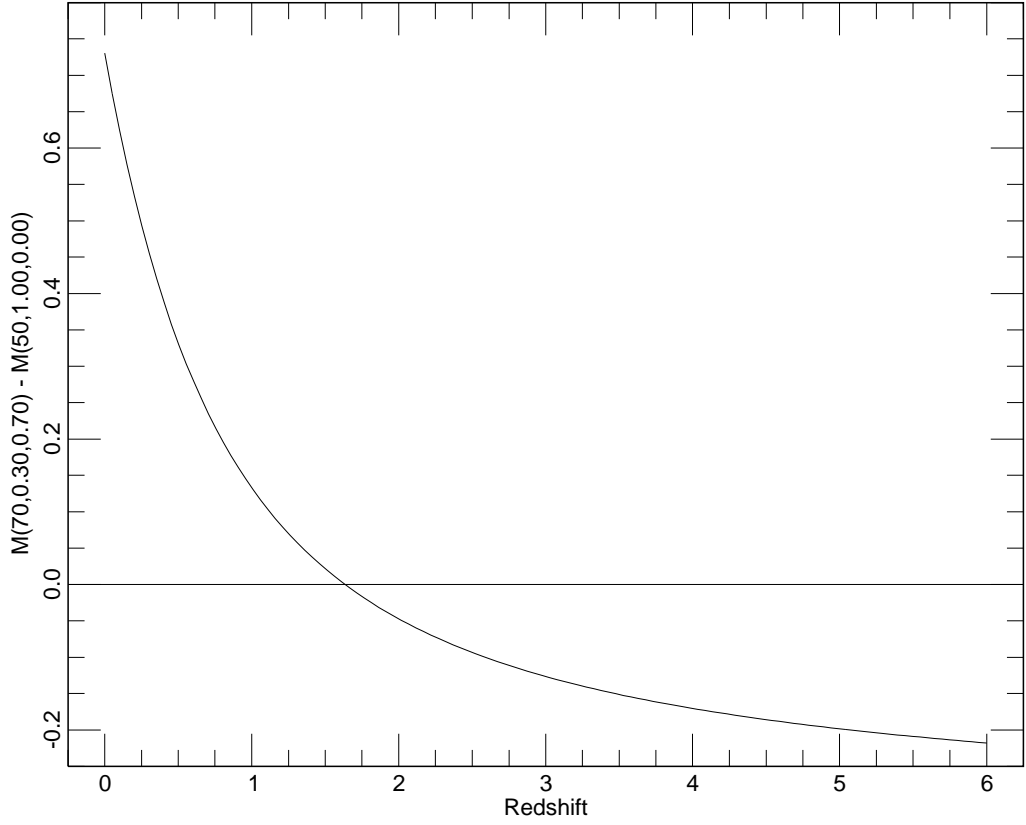}{4.5in}{90.0}{100.0}{100.0}{520.0}{0.0}
\figcaption{
The difference in absolute magnitude introduced by the change in
the adopted cosmology between the DR1 and EDR Quasar Catalogs.
The curved line in the figure is
\hbox{$M(H_0 = 70, \Omega_M = 0.3, \Omega_{\Lambda} = 0.7) -
M(H_0 = 50, \Omega_M = 1.0, \Omega_{\Lambda} = 0.0)$.}
At zero redshift the DR1 luminosities are lower than the EDR values
by~0.73~mag due to the change in Hubble's constant; the two cosmologies
produce equal luminosities at \hbox{$z \approx 1.7$,} and the DR1 luminosities
are about~20\% larger at redshifts above four.
\label{Figure 2 }
}
\end{figure}

\clearpage

\begin{figure}
\plotfiddle{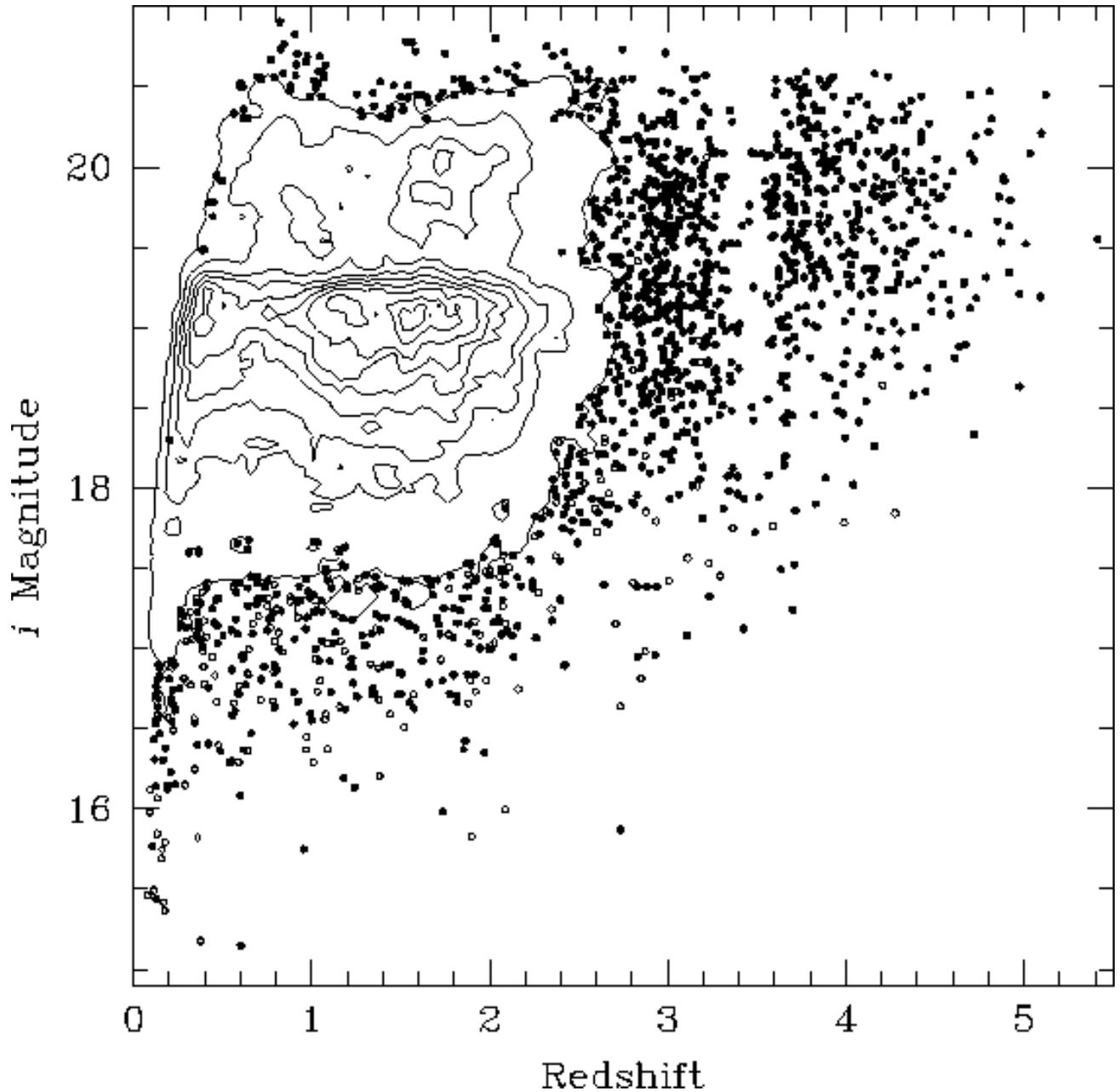}{6.5in}{0.0}{90.0}{90.0}{-270.0}{-100.0}
\figcaption{
The observed~$i$ magnitude as a function of redshift for the~16,713
objects in the catalog.  Open circles indicate quasars in NED that
were not discovered by
the SDSS.  Three quasars with \hbox{$i > 21$} are not plotted.
The distribution is represented by a set of linear contours when the
density of points in this two-dimensional space exceeds a certain threshold.
\label{Figure 3 }
}
\end{figure}

\clearpage

\begin{figure}
\plotfiddle{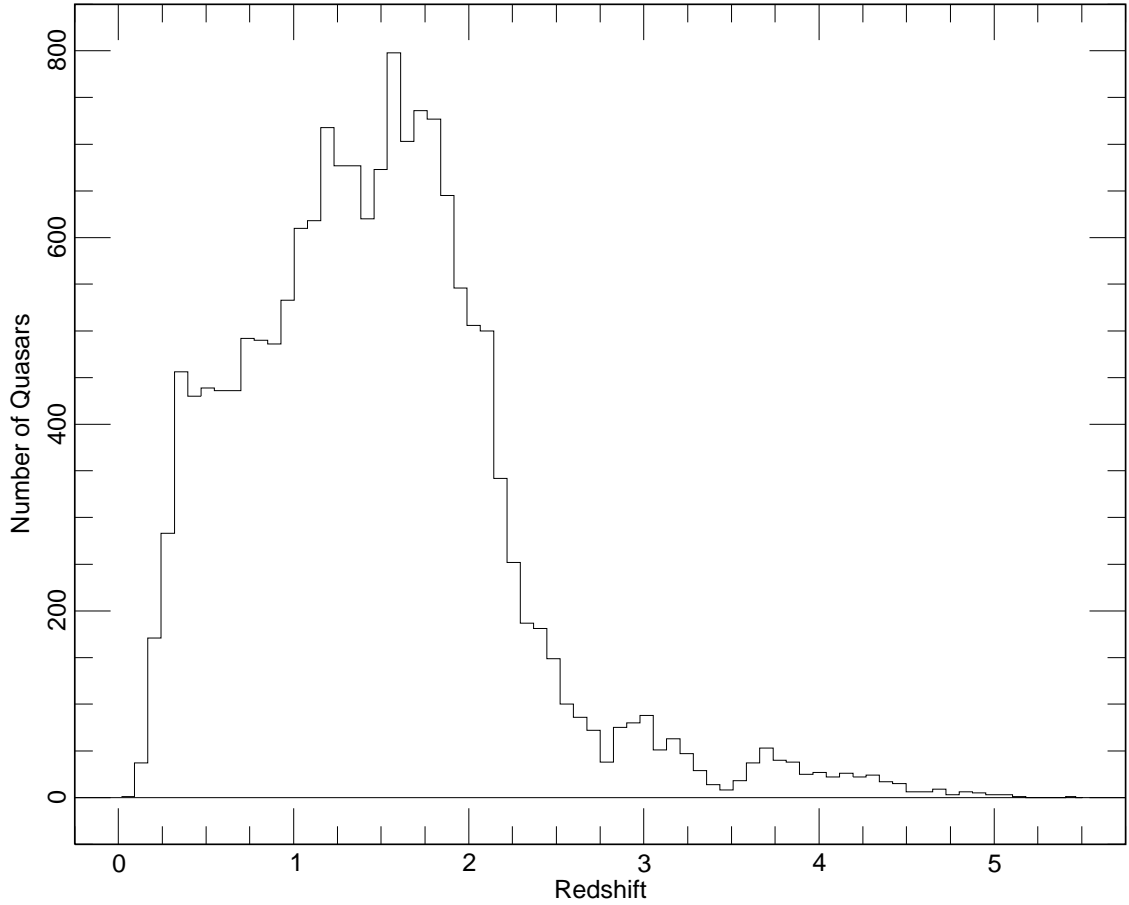}{5.5in}{90.0}{110.0}{110.0}{575.0}{0.0}
\figcaption{
The redshift histogram of the catalog quasars.  The smallest redshift is~0.08
and the largest redshift is~5.41; the median redshift of the catalog is~1.43.
The redshift bins have a width of~0.076.  The dips at redshifts of~2.7 and~3.5
are caused by the lower efficiency of the selection algorithm at these
redshifts.
\label{Figure 4 }
}
\end{figure}

\clearpage

\begin{figure}
\plotfiddle{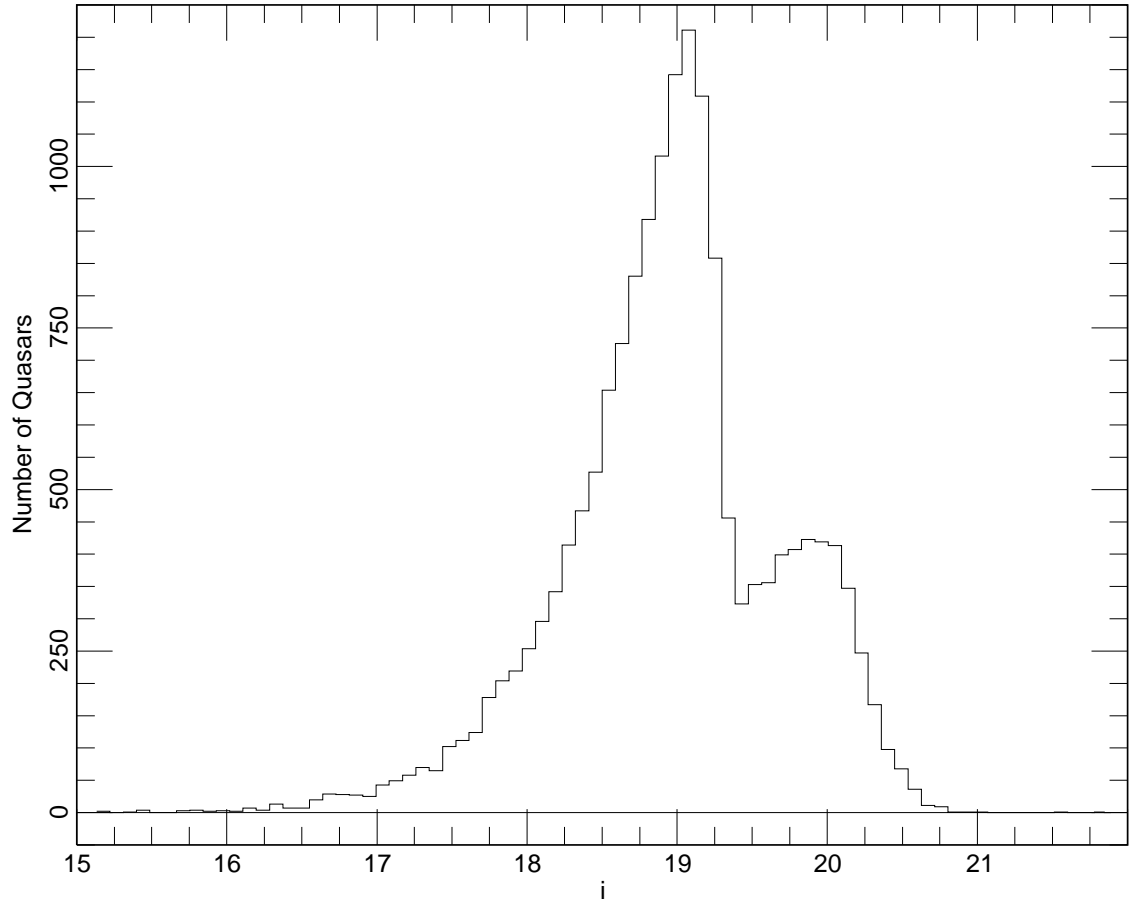}{5.5in}{90.0}{110.0}{110.0}{575.0}{0.0}
\figcaption{
The $i$ magnitude (not corrected for Galactic absorption) histogram of the
catalog quasars.  The magnitude bins have a width of~0.089.  The sharp
drop that occurs at magnitudes slightly fainter than 19 is due to the
low-redshift flux limit of the survey.
\label{Figure 5 }
}
\end{figure}

\clearpage

\begin{figure}
\plotfiddle{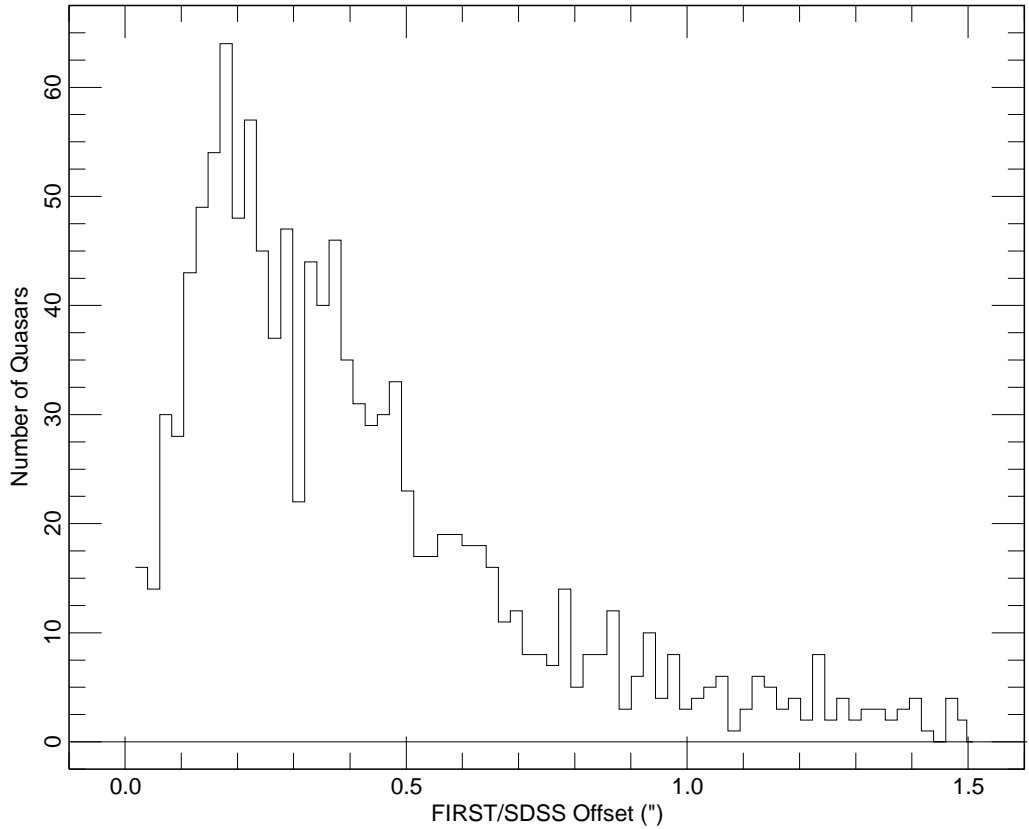}{5.0in}{90.0}{100.0}{100.0}{500.0}{0.0}
\figcaption{
Histogram of the offsets between the 1193 SDSS and FIRST matches; the
matching radius was set to~1.5$''$.  An analysis indicates that all of the
FIRST/SDSS matches are likely to be correct.
\label{Figure 6 }
}
\end{figure}

\clearpage

\begin{figure}
\plotfiddle{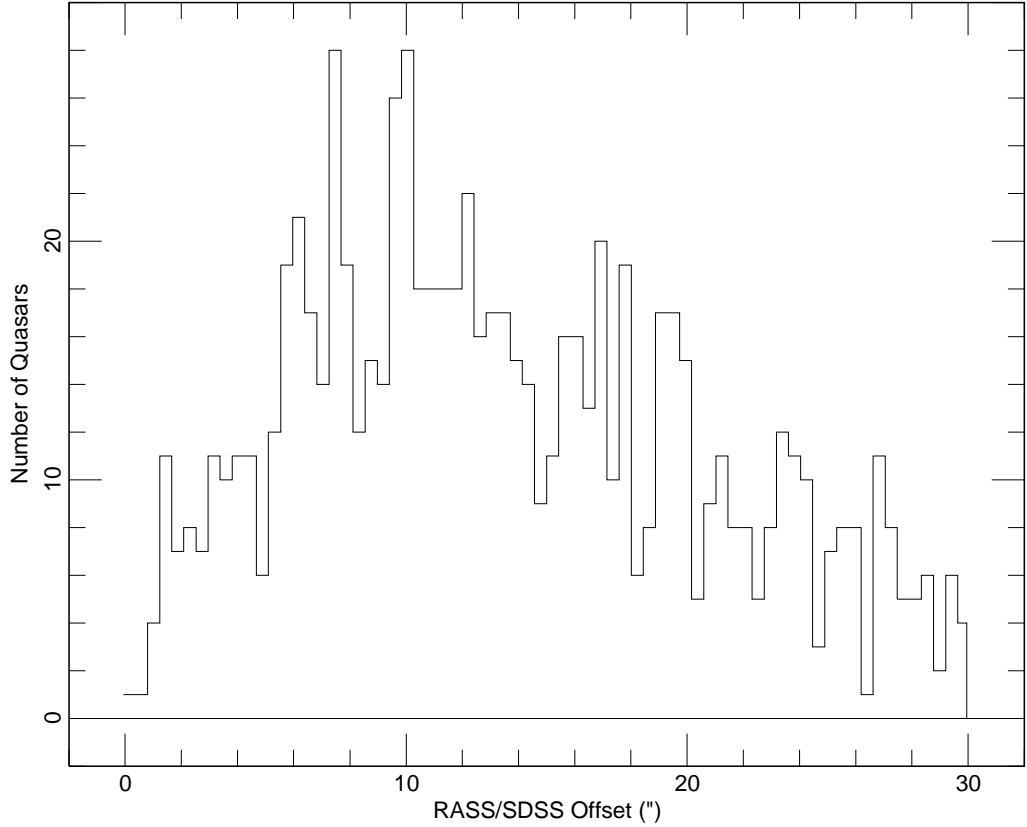}{5.0in}{90.0}{100.0}{100.0}{500.0}{0.0}
\figcaption{
Histogram of the offsets between the 824 SDSS and RASS FSC/BSC matches; the
matching radius was set to~30$''$. It is expected that 1-2\% of the X-ray
identifications are chance superpositions.
\label{Figure 7 }
}
\end{figure}

\clearpage

\begin{figure}
\plotfiddle{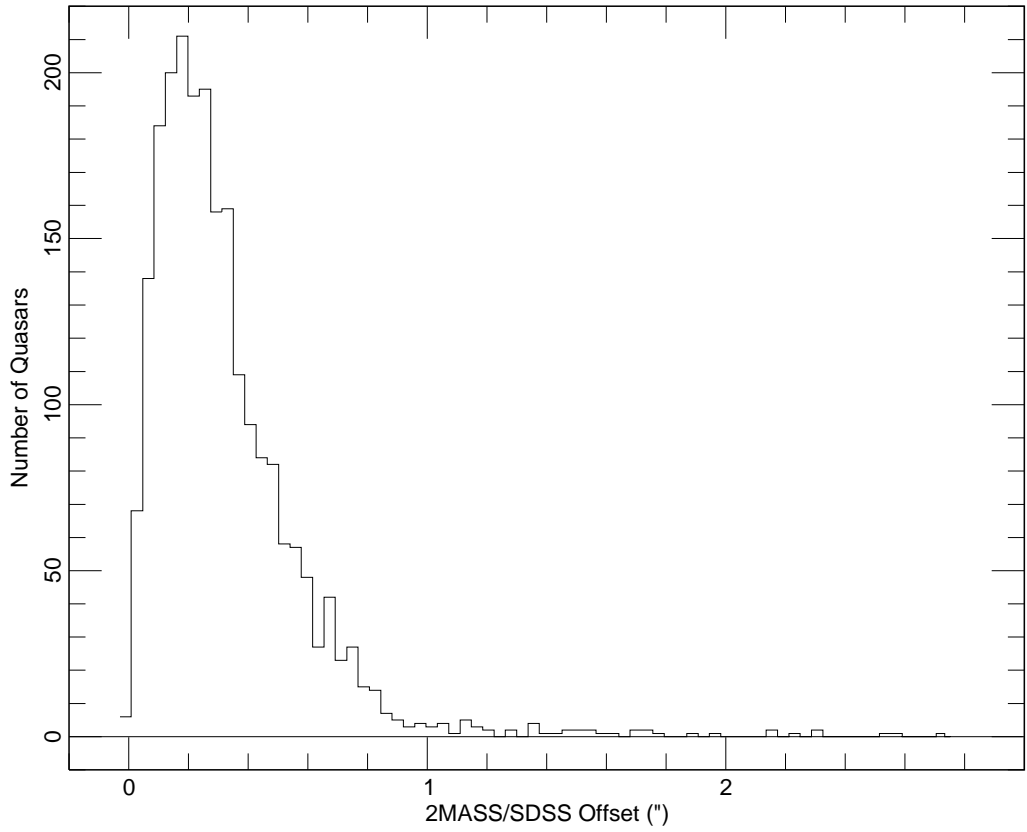}{5.0in}{90.0}{100.0}{100.0}{500.0}{0.0}
\figcaption{
Histogram of the offsets between the 2260 SDSS and 2MASS matches; the
matching radius was set to~3$''$.  Essentially all of these matches
are correct identifications.
\label{Figure 8 }
}
\end{figure}

\clearpage

\begin{figure}
\plotfiddle{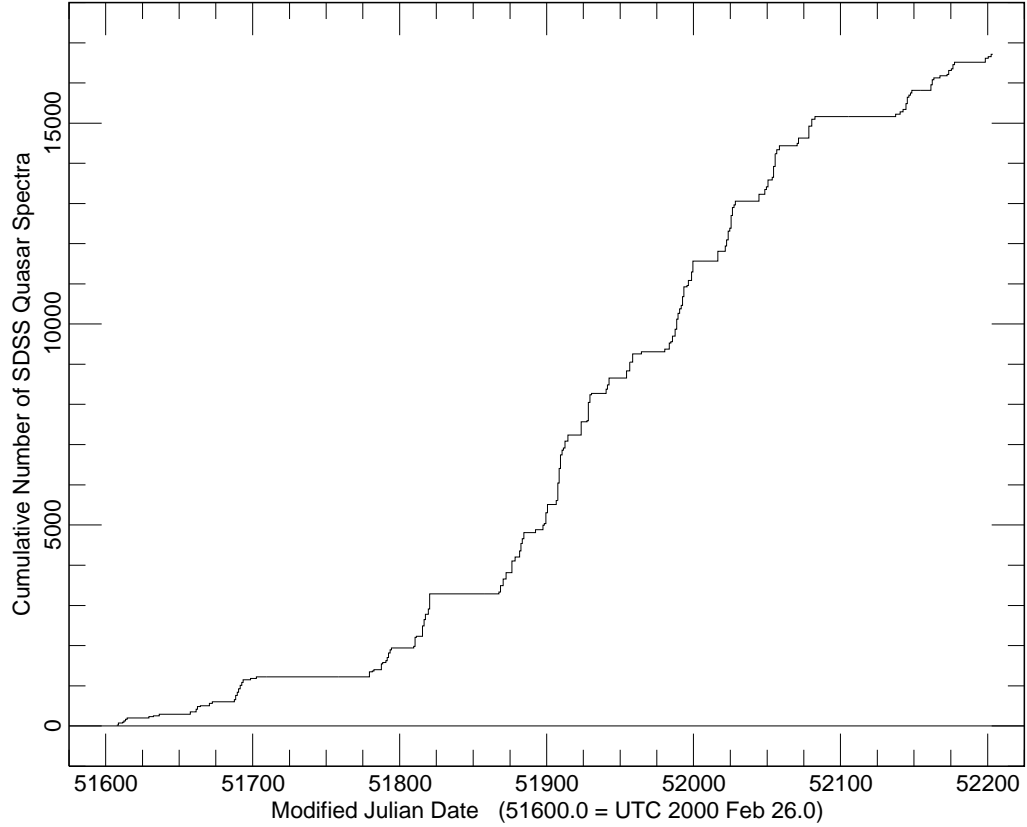}{5.0in}{90.0}{100.0}{100.0}{500.0}{0.0}
\figcaption{
The cumulative number of DR1 quasars as a function of time.  The horizonal
axis runs from February 2000 to October~2001.  The two long plateaus
reflect the SDSS summer shutdown periods.
\label{Figure 9 }
}
\end{figure}

\clearpage


\halign{\hskip 12pt
\hfil # \tabskip=1em plus1em minus1em&
\hfil # \hfil &
# \hfil \cr
\multispan3{\hfil TABLE 1 \hfil} \cr
\noalign{\medskip}
\multispan3{\hfil SDSS DR1 Quasar Catalog Format \hfil} \cr
\noalign{\bigskip\hrule\smallskip\hrule\medskip}
\hfil Column \hfil & \hfil Format \hfil & \hfil Description \hfil \cr
\noalign{\medskip\hrule\bigskip}
   1  &  A18  &   
SDSS DR1 Designation   \ \ \ \ hhmmss.ss+ddmmss.s  \ \ \ (J2000) \cr
   2  &  F11.6  &   Right Ascension in decimal degrees (J2000) \cr
   3  &  F11.6  &   Declination in decimal degrees (J2000) \cr
   4  &  F7.4 &   Redshift \cr
   5  &   I2  &   0 \ =  \ Pipeline Quasar \ \ \ \ \ 1 \ = \ Unusual Object
Search \cr
   6  &   F7.3  &    PSF $u$ magnitude
(not corrected for Galactic absorption) \cr
   7  &   F6.3  &    Error in PSF $u$ magnitude \cr
   8  &   F7.3  &    PSF $g$ magnitude
(not corrected for Galactic absorption) \cr
   9  &   F6.3  &    Error in PSF $g$ magnitude \cr
  10  &   F7.3  &    PSF $r$ magnitude 
(not corrected for Galactic absorption) \cr
  11  &   F6.3  &    Error in PSF $r$ magnitude \cr
  12  &   F7.3  &    PSF $i$ magnitude
(not corrected for Galactic absorption) \cr
  13  &   F6.3  &    Error in PSF $i$ magnitude \cr
  14  &   F7.3  &    PSF $z$ magnitude
(not corrected for Galactic absorption) \cr
  15  &   F6.3  &    Error in PSF $z$ magnitude \cr
  16  &   F7.3  &    Galactic absorption in $u$ band \cr
  17  &   F7.3  &    FIRST peak flux density at 20 cm expressed as AB magnitude;
\cr
& & \ \ \ \ \ 0.0 is no detection, $-1.0$ source is not in FIRST area \cr
  18  &   F8.3  &    S/N of FIRST flux density \cr
  19  &   F7.3  &    SDSS-FIRST separation in arc seconds \cr
  20  &   F8.3  &   log RASS full band count rate; 0.0 is no detection \cr
  21  &   F7.3  &   S/N of RASS count rate \cr
  22  &   F7.3  &   SDSS-RASS separation in arc seconds \cr
  23  &   F7.3  &   $J$ magnitude (2MASS);
0.0 indicates no 2MASS detection \cr
  24  &   F6.3  &   Error in $J$ magnitude (2MASS) \cr
  25  &   F7.3  &   $H$ magnitude (2MASS);
0.0 indicates no 2MASS detection \cr
  26  &   F6.3  &   Error in $H$ magnitude (2MASS) \cr
  27  &   F7.3  &   $K$ magnitude (2MASS);
0.0 indicates no 2MASS detection \cr
  28  &   F6.3  &   Error in $K$ magnitude (2MASS) \cr
  29  &   F7.3  &   SDSS-2MASS separation in arc seconds \cr
\noalign{\medskip\hrule}}

\clearpage

\halign{\hskip 12pt
\hfil # \tabskip=1em plus1em minus1em&
\hfil # \hfil &
# \hfil \cr
\multispan3{\hfil TABLE 1 \hfil} \cr
\noalign{\medskip}
\multispan3{\hfil SDSS DR1 Quasar Catalog Format (Continued) \hfil} \cr
\noalign{\bigskip\hrule\smallskip\hrule\medskip}
\hfil Column \hfil & \hfil Format \hfil & \hfil Description \hfil \cr
\noalign{\medskip\hrule\bigskip}
  30  &   F8.3  &   $M_{i}$ ($H_0$ = 70 km s$^{-1}$ Mpc$^{-1}$,
$\Omega_M = 0.3$, $\Omega_{\Lambda} = 0.7$, $\alpha = -0.5$ \cr
  31  &   I3  &   Morphology flag \ \ \ 0 = point source \ \ \ 1 = extended \cr
  32  &   I3  &   Quasar Target Selection Algorithm used for spectroscopy \cr
  33  &   I3  &   TARGET Spectroscopic Target flag: Low-$z$ Quasar
(0 or 1) \cr
  34  &   I3  &   TARGET Spectroscopic Target flag: High-$z$ Quasar
(0 or 1) \cr
  35  &   I3  &   TARGET Spectroscopic Target flag: FIRST (0 or 1) \cr
  36  &   I3  &   TARGET Spectroscopic Target flag: ROSAT (0 or 1) \cr
  37  &   I3  &   TARGET Spectroscopic Target flag: Serendipity (0 or 1) \cr
  38  &   I3  &   TARGET Spectroscopic Target flag: Star (0 or 1) \cr
  39  &   I3  &   TARGET Spectroscopic Target flag: Galaxy (0 or 1) \cr
  40  &   I3  &   BEST Spectroscopic Target flag: Low-$z$ Quasar (0 or 1) \cr
  41  &   I3  &   BEST Spectroscopic Target flag: High-$z$ Quasar (0 or 1) \cr
  42  &   I3  &   BEST Spectroscopic Target flag: FIRST (0 or 1) \cr
  43  &   I3  &   BEST Spectroscopic Target flag: ROSAT (0 or 1) \cr
  44  &   I3  &   BEST Spectroscopic Target flag: Serendipity (0 or 1) \cr
  45  &   I3  &   BEST Spectroscopic Target flag: Star (0 or 1) \cr
  46  &   I3  &   BEST Spectroscopic Target flag: Galaxy (0 or 1) \cr
  47  &   I6  &   SDSS Imaging Run Number of photometric measurements \cr
  48  &   I6  &   Modified Julian Date of imaging observation \cr
  49  &   I6  &   Modified Julian Date of spectroscopic observation \cr
  50  &   I5  &   Spectroscopic Plate Number \cr
  51  &   I5  &   Spectroscopic Fiber Number \cr
  52  &   1X, A25 &   Object Name for previously known quasars \cr
 & & \ \ \ ``SDSS" designates previously published SDSS object \cr
\noalign{\medskip\hrule}}

\clearpage

\halign{\hskip 12pt
\hfil $#$ \hfil \tabskip=1em plus1em minus1em&
\hfil # &
\hfil $#$ &
\hfil # &
\hfil # \hfil &
\hfil # &
\hfil # &
\hfil # &
\hfil # &
\hfil # &
\hfil # &
\hfil # &
\hfil # &
\hfil # &
\hfil # \cr
\multispan{15}{\hfil TABLE 2 \hfil} \cr
\noalign{\medskip}
\multispan{15}{\hfil The SDSS Quasar Catalog II\footnote{
The complete version of this table is in the electronic
edition of the Astronomical Journal and at {\tt 
http://www.sdss.org/dr1/products/value$\_$added/qsocat$\_$dr1.html}.  The catalog contains 52 columns of information
on 16,713 quasars.} \hfil} \cr
\noalign{\bigskip\hrule\smallskip\hrule\medskip}
}

\clearpage

\halign{\hskip 12pt
# \hfil \tabskip=1em plus1em minus1em&
\hfil # &
\hfil # &
\hfil # &
\hfil # \cr
\multispan5{\hfil TABLE 3 \hfil} \cr
\noalign{\medskip}
\multispan5{\hfil Spectroscopic Target Selection \hfil} \cr
\noalign{\bigskip\hrule\smallskip\hrule\medskip}
& \hfil TARGET \hfil & \hfil TARGET \hfil &
 \hfil BEST \hfil & \hfil BEST \hfil \cr
&&\hfil Sole \hfil && \hfil Sole \hfil \cr
\hfil Class \hfil & \hfil Selected \hfil & \hfil Selection \hfil &
\hfil Selected \hfil & \hfil Selection \hfil \cr
\noalign{\medskip\hrule\bigskip}
Low-$z$ & 11707 & 4322 & 10174 & 3263 \cr
High-$z$ & 2062 & 990 & 2717 & 428 \cr
FIRST  & 720 & 48 & 659 & 40 \cr
ROSAT  & 878 & 58 & 979 & 114 \cr
Serendipity & 10066 & 3399 & 9238 & 3637 \cr
Star & 932 & 42 & 150 & 37 \cr
Galaxy & 114 & 25 & 142 & 16 \cr
\noalign{\medskip\hrule}}

\clearpage
\halign{\hskip 12pt
# \hfil \tabskip=1em plus1em minus1em&
\hfil # \hfil &
\hfil $#$ \hfil &
# \hfil \cr
\multispan4{\hfil TABLE 4 \hfil} \cr
\noalign{\medskip}
\multispan4{\hfil Discrepant Redshifts \hfil} \cr
\noalign{\bigskip\hrule\smallskip\hrule\medskip}
\hfil Quasar (SDSS) \hfil & \hfil $z_{\rm SDSS}$ \hfil &
\hfil  z_{\rm NED} - z_{\rm SDSS} & NED Object Name/Notes \cr
\noalign{\medskip\hrule\bigskip}
J002411.65$-$004348.0 & 1.79 & -1.02 & LBQS 0021-0100            \cr
J004705.83$-$004819.5 & ... & ...   & From EDR; not a quasar \cr
J012428.10$-$001118.4 & 1.73 & -1.21 & SDSS                      \cr
J014905.28$-$011405.0 & 2.10 & -0.14 & SDSS                      \cr
J015032.88+143425.5 & 4.27 & -0.13 & SDSS                      \cr
\noalign{\smallskip}
J023044.81$-$004658.0 & 0.92 & +0.91 & SDSS                      \cr
J033305.32$-$053708.9 & 4.22 & -0.13 & SDSS                      \cr
J083223.22+491320.9 & 1.26 & -0.72 & [HB89] 0828+493           \cr
J084957.98+510829.0 & 0.58 & +1.28 & SBS 0846+513              \cr
J092004.31+591732.6 & 1.29 & -0.72 & SBS 0916+595              \cr
\noalign{\smallskip}
J093052.25+003458.8 & 1.77 & -1.26 & [HB89] 0928+008           \cr
J094031.07+551001.7 & 1.33 & +0.55 & SBS 0936+553B             \cr
J094443.08+580953.2 & 0.56 & +0.15 & SBS 0941+583              \cr
J095723.69+011458.7 & 2.48 & -1.56 & 2QZ J095723.6+011458      \cr
J101139.85+004039.4 & 1.71 & +0.60 & 2QZ J101139.8+004039      \cr
\noalign{\smallskip}
J101211.62+003719.3 & 1.63 & -0.89 & 2QZ J101211.6+003719      \cr
J114534.12+010308.0 & 1.07 & -0.67 & SDSS                      \cr
J115024.80+015620.3 & 0.71 & +0.80 & PMN J1150+0156            \cr
J120015.35+000553.1 & 1.65 & -1.28 & SDSS                      \cr
J120548.48+005343.8 & 0.93 & -0.83 & [HB89] 1203+011           \cr
\noalign{\smallskip}
J123113.95$-$021703.0 & 0.76 & -0.48 & [HB89] 1228-020           \cr
J133121.81+000248.4 & 3.22 & -2.35 & 2QZ J133121.8+000249      \cr
J140848.81+650528.0 & 1.94 & -0.93 & 87GB 140735.5+651947      \cr
J151307.26$-$000559.3 & 2.03 & -0.17 & SDSS                      \cr
J171528.24+550038.8 & 0.74 & -0.45 & SDSS                      \cr
\noalign{\smallskip}
J171930.24+584804.7 & 2.08 & -1.38 & SDSS                      \cr
J172542.16+582110.5 & ... & ... & Replaces EDR 172543.02+582110.8 \cr
J231324.45+003444.5 & 2.08 & +0.12 & H 2310+0018               \cr
\noalign{\medskip\hrule}}

\clearpage

\end{document}